\documentclass[twocolumn]{aastex63}
\usepackage[utf8]{inputenc}
\usepackage{makecell}
\usepackage{amsmath}
\usepackage{tikz}
\usepackage{hyperref}
\usepackage{soul}

\setstcolor{red}

\received{---}
\revised{---}
\accepted{---}

\shorttitle{
Outflows in Low-Mass Galaxies
}
\shortauthors{Xu et al.}

\begin{document}
\title{
EMPRESS. VI.\\
Outflows Investigated in Low-Mass Galaxies with $M_*=10^4-10^7~M_\odot$:\\ 
Weak Feedback in Low-Mass Galaxies?
}

\author[0000-0002-5768-8235]{Yi Xu}
\affiliation{Institute for Cosmic Ray Research, The University of Tokyo, 5-1-5 Kashiwanoha, Kashiwa, Chiba 277-8582, Japan}
\affiliation{Department of Astronomy, Graduate School of Science, the University of Tokyo, 7-3-1 Hongo, Bunkyo, Tokyo 113-0033, Japan}
\correspondingauthor{Yi Xu}
\email{xuyi@icrr.u-tokyo.ac.jp}

\author[0000-0002-1049-6658]{Masami Ouchi}
\affiliation{National Astronomical Observatory of Japan, 2-21-1 Osawa, Mitaka, Tokyo 181-8588, Japan}
\affiliation{Institute for Cosmic Ray Research, The University of Tokyo, 5-1-5 Kashiwanoha, Kashiwa, Chiba 277-8582, Japan}
\affiliation{Kavli Institute for the Physics and Mathematics of the Universe (Kavli IPMU, WPI), The University of Tokyo, 
Kashiwanoha 5-1-5, Kashiwa, Chiba 277-8583, Japan}

\author{Michael Rauch}
\affiliation{Carnegie Observatories, 813 Santa Barbara Street, Pasadena, CA 91101, USA}

\author[0000-0003-2965-5070]{Kimihiko Nakajima}
\affiliation{National Astronomical Observatory of Japan, 2-21-1 Osawa, Mitaka, Tokyo 181-8588, Japan}

\author[0000-0002-6047-430X]{Yuichi Harikane}
\affiliation{Institute for Cosmic Ray Research, The University of Tokyo, 5-1-5 Kashiwanoha, Kashiwa, Chiba 277-8582, Japan}
\affiliation{Department of Physics and Astronomy, University College London, Gower Street, London WC1E 6BT, UK}

\author[0000-0001-6958-7856]{Yuma Sugahara}
\affiliation{National Astronomical Observatory of Japan, 2-21-1 Osawa, Mitaka, Tokyo 181-8588, Japan}
\affiliation{Waseda Research Institute for Science and Engineering, Faculty of Science and Engineering, Waseda University, 3-4-1, Okubo, Shinjuku, Tokyo 169-8555, Japan}

\author[0000-0002-3852-6329]{Yutaka Komiyama}
\affiliation{National Astronomical Observatory of Japan,
2-21-1 Osawa, Mitaka, Tokyo 181-8588, Japan}
\affiliation{Graduate University for Advanced Studies (SOKENDAI), 2-21-1 Osawa, Mitaka, Tokyo 181-8588, Japan}

\author[0000-0002-3801-434X]{Haruka Kusakabe}
\affiliation{Observatoire de Gen\`{e}ve, Universit\'e de Gen\`{e}ve, 51 Chemin de P\'egase, 1290 Versoix, Switzerland}

\author[0000-0001-7201-5066]{Seiji Fujimoto}
\affiliation{Cosmic Dawn Center (DAWN), Jagtvej 128, DK2200 Copenhagen N, Denmark}
\affiliation{Niels Bohr Institute, University of Copenhagen, Lyngbyvej 2, DK2100 Copenhagen \O, Denmark}

\author[0000-0001-7730-8634]{Yuki Isobe}
\affiliation{Institute for Cosmic Ray Research, The University of Tokyo, 5-1-5 Kashiwanoha, Kashiwa, Chiba 277-8582, Japan}
\affiliation{Department of Physics, Graduate School of Science, The University of Tokyo, 7-3-1 Hongo, Bunkyo, Tokyo 113-0033, Japan}

\author[0000-0002-1418-3309]{Ji Hoon Kim}
\affiliation{Subaru Telescope, National Astronomical Observatory of Japan, National Institutes of Natural Sciences (NINS), 650 North Aohoku Place, Hilo, HI 96720, USA}
\affiliation{Metaspace, 36 Nonhyeon-ro, Gangnam-gu, Seoul 06312, Republic of Korea}

\author[0000-0001-9011-7605]{Yoshiaki Ono}
\affiliation{Institute for Cosmic Ray Research, The University of Tokyo, 5-1-5 Kashiwanoha, Kashiwa, Chiba 277-8582, Japan}

\author[0000-0001-7869-2551]{Fakhri S. Zahedy}
\affiliation{Carnegie Observatories, 813 Santa Barbara Street, Pasadena, CA 91101, USA}

\begin{abstract}
We study emission line profiles of 21 nearby low-mass ($M_*=10^4-10^7~M_\odot$) galaxies in deep medium-high resolution spectra taken with Magellan/MagE.
These low-mass galaxies are actively star-forming systems with high specific star-formation rates of $\mathrm{sSFR}\sim100-1000~\mathrm{Gyr}^{-1}$ that are well above the star-formation main sequence and its extrapolation.
We identify broad-line components of H$\alpha$ and {\sc [Oiii]}$\lambda 5007$ emission in 14 out of the 21 galaxies that cannot be explained by the MagE instrumental profile or the natural broadening of line emission.
We conduct double Gaussian profile fitting to the emission of the 14 galaxies, and find that the broad-line components have line widths significantly larger than those of the narrow-line components, indicative of galactic outflows.
The board-line components have moderately large line widths of $\sim 100$ km s$^{-1}$. 
We estimate the maximum outflow velocities $v_\mathrm{max}$ and obtain values of $\simeq 60-200$ km s$^{-1}$, which are found to be comparable to or slightly larger than the escape velocities.
Positive correlations of $v_\mathrm{max}$ with star-formation rates, stellar masses, and circular velocities, extend down into this low-mass regime.
Broad- to narrow-line flux ratios BNRs are generally found to be smaller than those of massive galaxies.
The small $v_\mathrm{max}$ and BNRs suggest that the mass loading factors $\eta$ can be as small as 0.1 - 1 or below, in contrast to the large $\eta$ of energy-driven outflows predicted by numerical simulations.
\end{abstract}

\keywords{galaxies: dwarf --- galaxies: evolution --- galaxies: kinematics and dynamics}

\section{Introduction}
Star formation is expected to have a strong impact on the interstellar medium (ISM).
In particular, young massive stars have strong stellar winds and radiation pressure capable of expelling the surrounding gas.
At a later stage, the massive stars create supernova explosions that heat the surrounding gas through thermal energy input.
These interactions play an important role in the evolution of galaxies.
The lack of cold gas leads to the suppresses of star formation activity, which is known as stellar feedback \citep{Cole+00,Hopkins+12,Leroy+15}.
Star formation also contributes to the metal enrichment of the intergalactic medium by ejecting gas from galaxies \citep{Tremonti+04,Dayal+13,Andrews&Martini13}.
In particular, stellar feedback is effective in galaxies with low stellar masses ($M_*$) whose shallow gravitational potential wells may not  be able to retain the gas.
The shallow potential may also be the cause of the increasing fraction of escaping gas from $M_*\sim10^{12}~M_\odot$ towards $M_*\sim10^9~M_\odot$ \citep{Arribas+14,Bruno+19}.
Moreover, stellar feedback may be able to explain two well-known observational results for low-mass galaxies that differ from the predictions of dark matter only numerical simulations, 
the observed number of low-mass galaxies and the density profile of DM halo \citep{Brooks19}.

Galactic outflow are among the most important and noticeable results of stellar feedback.
The strength of stellar feedback can be quantified by the mass-loading factor ($\eta$), which is defined as the ratio of the mass loss rate ($\dot{M}_\mathrm{out}$) to the star formation rate (SFR).
Semianalytic models have adopted prescriptions for the feedback process assuming either energy-driven outflows with $\eta\propto v_\mathrm{cir}^{-2}$ or momentum-driven outflows with $\eta\propto v_\mathrm{cir}^{-1}$ (see Equations 13 and 35 in \citealt{Murray+05}), where $v_\mathrm{cir}$ is the circular velocity of the DM halo.
Recent simulations can resolve the star formation activity and the outflowing gas in individual galaxies.
However, the ``sub-grid'' physics of outflows still relies on theoretical or empirical formulae  \citep[e.g.,][]{Muratov+15,Christensen+18,Hu19}.
In this paper we will focus on low-mass galaxies below $M_*\sim10^7~M_\odot$, a mass range for which the ``sub-grid'' physics is particular poorly understood, and which has not been fully explored by previous observational studies.

It is known that the outflows are composed of hot ($T\sim10^6~\mathrm{K}$), warm ($T\sim10^4~\mathrm{K}$), and cold phases ($T\sim10^1-10^3~\mathrm{K}$) \citep[e.g.,][]{Veilleux05}.
In the present work, we focus on the warm-phase gas that is found to be widely extended in young low-mass star-forming galaxies \citep{Pardy+16}.
For warm-phase gas, various methods are used in the literature to measure gas motion with emission lines, including two-component fitting \citep[e.g.,][]{Freeman+19,Bruno+19}, 
non-paramatric procedures \citep[e.g.,][]{Cicone+16},
and spatially resolved line mapping \citep[e.g.,][]{McQuinn+19,Bik+15}.
The two-component fitting decomposes the emission profile into narrow and broad components.
The narrow components are broadened by the rotation or dispersion of gas, while the broad components are supposed to trace the motion of outflowing gas.
The two-component fitting of emission lines requires
high spectral resolution to resolve the emission line profiles.
Observations are particularly challenging for low-mass galaxies, due to low brightness and intrinsically narrow emission.

This paper is the sixth paper of a program named ``Extremely Metal-Poor Representatives Explored by the Subaru Survey (EMPRESS)'' started by \citeauthor{Kojima+20a} (\citeyear{Kojima+20a}; hereafter K20). 
Using machine-learning techniques, K20 select extremely metal-poor galaxies (EMPGs) from their source catalogs that are constructed with the data from Subaru/Hyper Suprime-Cam Subaru Strategic Program (HSC-SSP, \citealt{HSC}) and the 13th release of Sloan Digital Sky Survey (SDSS DR13, \citealt{SDSS13}).
K20 identify 113 EMPG candidates in the local universe, that have compact sizes and possibly very small stellar masses.
This paper describes a dataset with sufficient spectral resolution and signal-to-noise ratio (SNR) to study the properties of such objects.
In Section \ref{observations}, we describe our MagE observations and the data reduction. 
In Section \ref{sample}, we present our galaxy sample, and in Section \ref{analysis}, we fit the spectra and derive galaxy properties. In Section \ref{results}, we present our results on outflow properties.
In Sections \ref{discussion} and \ref{summary}, we discuss and summarize our observational results, respectively.
Throughout the paper we adopt a cosmological model with $H_0=70~\mathrm{km~s^{-1}~Mpc^{-1}}$, $\Omega_\Lambda=0.7$, and $\Omega_{m}=0.3$.

\begin{deluxetable}{ccccc}
    \label{tab:obsobj}
    \tablecaption{Summary of the MagE Observations}
    \tablewidth{0pt}
    \tablehead{
        \colhead{ID} & \colhead{R.A.} & \colhead{Decl.} & \colhead{Exposure} & \colhead{Ref.} \\
        \colhead{} & \colhead{(hh:mm:ss)} & \colhead{(dd:mm:ss)} & \colhead{(sec)} & \colhead{} \\
        \colhead{(1)} & \colhead{(2)} & \colhead{(3)} & \colhead{(4)} & 
        \colhead{(5)}
    }
    \startdata
    \multicolumn{5}{c}{Photometric Candidates}\\
    \hline
    J0845$+$0131 & 08:45:30.81 & $+$01:31:51.19 & 1,800 & (a)\\
    J0912$-$0104 & 09:12:18.12 & $-$01:04:18.32 & 2,700 & (a)\\
    J0935$-$0115 & 09:35:39.20 & $-$01:15:41.41 & 1,350 & (a)\\
    J1011$+$0201 & 10:11:47.80 & $+$02:01:08.30 & 1,400 & (a)\\
    J1210$-$0103 & 12:10:33.54 & $-$01:03:11.69 & 2,700 & (a)\\
    J1237$-$0016 & 12:37:47.89 & $-$00:16:00.76 & 1,800 & (a)\\
    J1401$-$0040 & 14:01:07.61 & $-$00:40:50.10 & 1,800 & (a)\\
    J1411$-$0032 & 14:11:03.69 & $-$00:32:40.77 & 1,800 & (a)\\
    J1452$+$0241 & 14:52:55.28 & $+$02:41:01.31 & 1,800 &(b)\\
    J1407$-$0047 & 14:07:10.69 & $-$00:47:26.31 & 1,800 &(b)\\
    \hline\hline
    \multicolumn{5}{c}{Spectroscopically-Confirmed Galaxies}\\
    \hline
    J1044$+$0353 & 10:44:57.80 & $+$03:53:13.30 & 900 & (c)\\
    J1253$-$0312 & 12:53:05.97 & $-$03:12:58.49 & 600 & (d)\\
    J1323$-$0132 & 13:23:47.46 & $-$01:32:51.94 & 1,800 & (e)\\
    J1418$+$2102 & 14:18:51.13 & $+$21:02:40.02 & 1,200 & (f)\\
    \enddata
    \tablecomments{Columns: (1) ID. (2) R.A. in J2000. (3) Declination in J2000. (4) Total exposure time. (5) Reference. (a) \cite{Kojima+20a}, (b) K. Nakajima et al. in preparation, (c) \cite{Kniazev+03}, (d) \cite{Kniazev+04}, (e) \cite{Izotov+12}, and (f) \cite{Sanchez-Almeida+16}.}
\end{deluxetable}

\section{Observations and Data Reduction}
\label{observations}
We carry out deep spectroscopy for 14 targets, 10 out of which are EMPG candidates from K20 and K. Nakajima et al. (in preparation).
The others, 4 targets, are spectroscopically-confirmed bright local dwarf galaxies taken from \cite{Kniazev+03}, \cite{Kniazev+04}, \cite{Izotov+12}, and \cite{Sanchez-Almeida+16}.
Our targets are summarized in Table \ref{tab:obsobj}.

\subsection{Observations}
All the targets were observed with the Magellan Echellette (MagE) spectrograph mounted on the Magellan Baade Telescope on 2021 February 9th (PI: M. Rauch). 
A $0''.70 \times10''$ slit was used at the parallactic angle to avoid wavelength-dependent slit-losses.
% We placed the slit with each target at the central position.
Although some targets have the feature of a bright clump on a diffused tail that is known as the tadpole morphology \cite[e.g.,][]{Morales-Luis+11,SanchezAlmeida+13},
the observing program was aimed at exploring the bright clumps.
Thereofre, we placed the slit to cover the bright clumps.
For each target, we obtained 2-3 science frames with exposure times of $300-900$ seconds depending on the luminosity of the target.
We took Qh lamp and in-focus (out-of-focus) Xe flash frames as the flat-fielding data for red and blue (very blue) orders of spectra, respectively.
The standard stars, HD49798 and CD329927, were observed at the beginning and the end of the observing run, respectively.
During the observations, the sky was clear with a typical seeing of $\sim0''.6$.

\subsection{Data Reduction}
We reduce the MagE spectroscopic data with PypeIt \citep{pypeit:joss_arXiv,pypeit:zenodo}, an open-source reduction package for spectroscopic data written in Python.
Our data reduction basically follows standard procedures, including flat fielding, wavelength calibration, sky subtraction, cosmic ray removal, and extracting one-dimensional (1D) spectra. 
We separately use two types of frames, the Qh lamp and Xe flash, for flat-fielding, and find no differences in the flat-fielded frames of 1D spectra in blue orders. 
However, Xe flash frames present broad emission line features in red orders. 
Thus, we choose the flat-fielding results given with Qh lamp frames from blue to red orders.

PypeIt does not fit a model of sky background correctly around bright extended emission lines (e.g. H$\alpha$ and [{\sc Oiii}]$\lambda 5007$). 
Given the fact that most emission lines we investigate do not overlap with strong skylines, we adopt flat sky background in pixels where strong emission lines exist and estimate the flux of sky background by averaging sky background in the nearby pixels.
We perform boxcar extraction using the sky subtracted 2D spectra.
We find no clear emission lines for J1011$+$0201.
For the other 13 targets, PypeIt successfully identifies the spatial position on each 2D spectrum that corresponds to the bright clump of the target, while the emission from the diffused tail is not clearly seen.
Around the position identified by PypeIt, a spatial extraction width is manually selected to extract a 1D spectrum that well represents the bright clump.
The 1D spectra extracted from the different exposures of each target are combined via PypeIt using \texttt{pypeit\_coadd\_1dspec}.
The coadded 1D spectra have a spectral resolution of $\sim27.8~\mathrm{km~s^{-1}}$ that is estimated from the unresolved emission lines of the lamp data (see Section \ref{fitting1d}).
Flux errors composed of read-out and photon noise are estimated and propagated to 1D spectra by PypeIt.
Of the two standard stars, HD49798 and CD329927, we choose whichever was closer to each target in declination for flux calibration.
We also apply telluric corrections via PypeIt using \texttt{pypeit\_tellfit}.

\begin{figure}[ht!]
    \centering
    \includegraphics[width=\linewidth]{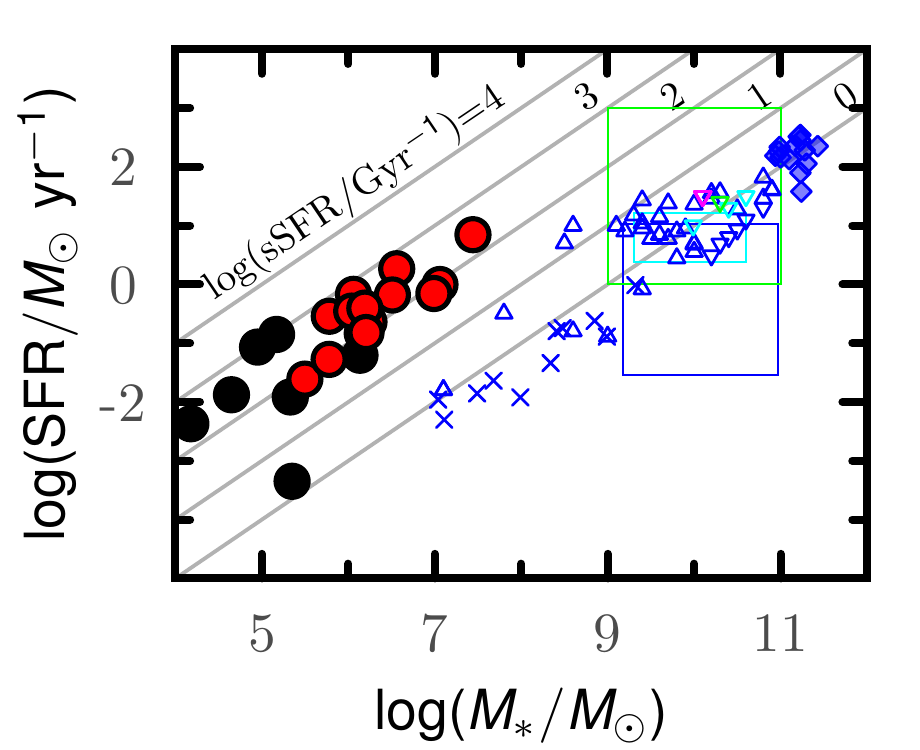}
    \caption{Stellar masses and SFRs.
    The red and black circles are our sample.
    For the galaxies represented by red (black) circles, the emission lines have double-Gaussian (single-Gaussian) profiles.
    The blue, cyan, green, and magenta symbols indicate galaxies at $z\sim0, 1, 2$, and $5-6$ that we take from previous outflow studies.
    For those we cannot obtain data for individual galaxies, parameter ranges are shown as rectangles.
    The blue rectangle, diamonds, and crosses represent galaxies investigated by \cite{Bruno+19}, \cite{Perrotta+21}, and \cite{McQuinn+19}, respectively, who use emission lines to study warm-phase outflows.
    Note that \cite{Bruno+19} and \cite{Perrotta+21} conduct double-Gaussian profile fitting to the emission lines, while \cite{McQuinn+19} uses single-Gaussian fitting.
    The cyan and green rectangles represent galaxies studied by \cite{Swinbank+19} and \cite{Freeman+19}, respectively, who use stacked H$\alpha$ emission.
    The blue open triangles are z$\sim$0 galaxies investigated by \cite{Heckman+15} and \cite{Heckman+16} using absorption line methods. 
    The blue, cyan, green, and magenta inverse triangles are taken from \cite{Sugahara+17,Sugahara+19} who investigate galaxies at different redshifts with absorption line methods.
    The grey lines indicate specific star formation rates of $\log(\mathrm{sSFR/Gyr^{-1}})=(0, 1, 2, 3, 4)$.}
    \label{fig:sfr_m}
\end{figure}

\begin{deluxetable*}{cCCCCCCCC}
    % \tablenum{2}
    \tablecaption{Our Sample}
    \tablewidth{0pt}
    \tablehead{
        \colhead{ID} & \colhead{redshift} & \colhead{$F(\mathrm{H}\beta)$} &
        % \colhead{EW$_0(\mathrm{H}\beta)$} & \colhead{$12+\log(\mathrm{O/H})$} &
        \colhead{$\log(M_*)$} & \colhead{$\log(\mathrm{SFR})$} &
        \colhead{$E(B-V)$} & \colhead{age} &
        \colhead{$M_\mathrm{h}$} & \colhead{$v_\mathrm{cir}$}\\
        \colhead{} & \colhead{} & \colhead{($\mathrm{erg~s^{-1}~cm^{-2}}$)} &
        % \colhead{(\AA)} & \colhead{} & 
        \colhead{($M_\odot$)} &
        \colhead{($M_\odot~\mathrm{yr^{-1}}$)} & \colhead{(mag)} &
        \colhead{(Myr)} &
        \colhead{($M_\odot$)} & \colhead{($\mathrm{km~s^{-1}}$)}\\
        \colhead{(1)} & \colhead{(2)} & \colhead{(3)} &
        \colhead{(4)} & \colhead{(5)} & \colhead{(6)} &
        \colhead{(7)} & \colhead{(8)} &
        \colhead{(9)}
    }
    \startdata
    % \cutinhead{Data from K20}
    \multicolumn{9}{c}{Data from K20}\\
    \hline
    J0002$+$1715 & 0.02083 & 45.7\pm1.4 & 7.06\pm0.03 & -0.002\pm0.013 & 0.00^{+0.01}_{-0.00} & 31 & 10.18\pm0.25 & 28.68^{+5.98}_{-4.95}\\
    J1142$-$0035 & 0.02035 & 4.27\pm0.15 & 4.95^{+0.04}_{-0.01} & -1.066\pm0.013 & 0.00^{+0.02}_{-0.00} & 3.7 & 9.50^{+0.25}_{-0.24} & 17.01^{+3.60}_{-2.87}\\
    J1642$+$2233 & 0.01725 & 46.3\pm1.5 & 6.06^{+0.03}_{-0.13} & -0.169\pm0.013 & 0.02\pm0.02 & 25 & 9.86^{+0.25}_{-0.28} & 22.39^{+4.67}_{-4.32}\\
    J2115$-$1734 & 0.02296 & 69.9\pm2.1 & 6.56\pm0.02 & 0.266\pm0.013 & 0.17\pm0.04 & 21 & 10.02\pm0.24 & 25.34^{+5.21}_{-4.32}\\
    J2253$+$1116 & 0.00730 & 139\pm4.28 & 5.78\pm0.01 & -0.541\pm0.013 & 0.00^{+0.01}_{-0.00} & 4.1 & 9.76\pm0.24 & 20.89^{+4.23}_{-3.52}\\
    J2310$-$0211 & 0.01245 & 99.3\pm3.1 & 6.99\pm0.03 & -0.155\pm0.013 & 0.01^{+0.02}_{-0.01} & 51 & 10.16\pm0.25 & 28.18^{+5.88}_{-4.87}\\
    J2314$-$0154 & 0.03265 & 2.61\pm0.10 & 5.17\pm0.01 & -0.851\pm0.013 & 0.28\pm0.03 & 4.1 & 9.57\pm0.24 & 17.96^{+3.64}_{-3.03}\\
    J2327$-$0200 & 0.01812 & 40.7\pm1.2 & 6.51^{+0.02}_{-0.03} & -0.180\pm0.013 & 0.00^{+0.02}_{-0.00} & 22 & 10.00^{+0.24}_{-0.25} & 25.02^{+5.15}_{-4.32}\\
    \hline
    \multicolumn{9}{c}{Data from this Study}\\
    \hline
    J0845$+$0131 & 0.01333 & 2.90\pm0.07 & 4.65^{+0.10}_{-0.08} & -1.88^{+0.08}_{-0.09} & 0.23\pm0.02 & 3.6 & 9.40^{+0.27}_{-0.26} & 15.78^{+3.63}_{-2.88}\\
    J0912$-$0104 & 0.02737 & 0.59\pm0.03 & 5.33^{+0.15}_{-0.27} & -1.91\pm0.04 & 0.08\pm0.04 & 12 & 9.62^{+0.28}_{-0.32} & 18.68^{+4.58}_{-4.12}\\
    J0935$-$0115 & 0.01621 & 22.18\pm0.57 & 6.16^{+0.16}_{-0.13} & -0.83^{+0.07}_{-0.08} & 0.18\pm0.02 & 18 & 9.89^{+0.29}_{-0.28} & 22.27^{+4.89}_{-4.00}\\
    J1044$+$0353 & 0.01317 & 77.34\pm1.90 & 6.04^{+0.07}_{-0.06} & -0.45\pm0.08 & 0.20\pm0.02 & 3.2 & 9.85\pm0.26 & 22.31^{+4.92}_{-4.09}\\
    J1210$-$0103 & 0.00865 & 0.17\pm0.03 & 5.35^{+0.14}_{-0.16} & -3.35^{+0.12}_{-0.13} & 0^{+0.20}_{-0.00} & 261 & 9.63^{+0.28}_{-0.29} & 18.78^{+4.54}_{-3.74}\\
    J1237$-$0016 & 0.05045 & 3.04\pm0.06 & 6.24\pm0.03 & -0.64\pm0.03 & 0.30\pm0.01 & 2.9 & 9.91\pm0.25 & 23.38^{+4.96}_{-4.06}\\
    J1253$-$0312 & 0.02301 & 467.96\pm16.21 & 7.44\pm0.05 & 0.84^{+0.06}_{-0.07} & 0.24\pm0.02 & 4.7 & 10.30\pm0.25 & 31.51^{+6.78}_{-5.60}\\
    J1323$-$0132 & 0.02275 & 26.41\pm5.90 & 6.19\pm0.05 & -0.40^{+0.05}_{-0.06} & 0.20\pm0.01 & 3.3 & 9.90\pm0.25 & 23.13^{+4.99}_{-4.07}\\
    J1401$-$0040 & 0.01168 & 6.31\pm0.11 & 5.50^{+0.08}_{-0.09} & -1.61^{+0.08}_{-0.09} & 0.20\pm0.01 & 4.2 & 9.67^{+0.26}_{-0.27} & 19.48^{+4.38}_{-3.60}\\
    J1407$-$0047 & 0.05368 & 0.73\pm0.02 & 6.13^{+0.07}_{-0.16} & -1.20\pm0.03 & 0.21\pm0.01 & 28 & 9.88^{+0.26}_{-0.29} & 22.80^{+5.05}_{-4.55}\\
    J1411$-$0032 & 0.02617 & 2.48\pm0.07 & 5.78^{+0.06}_{-0.05} & -1.27\pm0.04 & 0.17\pm0.01 & $<1$ & 9.76^{+0.26}_{-0.25} & 20.88^{+4.54}_{-3.70}\\
    J1418$+$2102 & 0.00889 & 68.82\pm74.84 & 6.20^{+0.25}_{-0.15} & -0.82^{+0.10}_{-0.11} & 0.17\pm0.01 & 8.2 & 9.90^{+0.32}_{-0.29} & 23.20^{+6.43}_{-4.58}\\
    J1452$+$0241 & 0.00574 & 4.62\pm0.09 & 4.18^{+0.13}_{-0.16} & -2.37^{+0.15}_{-0.18} & 0.26\pm0.02 & 3.3 & 9.25^{+0.28}_{-0.29} & 14.05^{+3.35}_{-2.80}
    \enddata
    \tablecomments{Columns: (1) ID. (2) Redshift. (3) H$\beta$ emission line flux normalized in units of $10^{-15}\mathrm{erg~s^{-1}~cm^{-2}}$. The fluxes are corrected for dust extinction. (4) Stellar mass. (5) Star-formation rate. (6) Color excess. (7) Maximum stellar age. (8) DM Halo mass. (9) circular velocity.}
    % \end{tabular}
    \label{tab:sample}
\end{deluxetable*}

\section{Sample and spectroscopic data}
\label{sample}
Our sample in this paper is composed of two sets of MagE spectra obtained in this study (Section \ref{observations}) and K20.
Exploiting the high spectral resolution of MagE, we try to resolve spectral profiles of emission lines in our sample.

\subsection{Data from this Study}
In this study, ten EMPG candidates from K20 and K. Nakajima et al. (in preparation) are observed with MagE (Section \ref{observations}). 
We identify that nine targets out of the ten EMPG photometric candidates are galaxies with confirmed emission lines, while one object is a Galactic star. 
We also investigate the MagE spectra of 4 bright spectroscopically-confirmed EMPGs (Section \ref{observations}), and find that S/Ns of their strong emission lines are sufficiently high.
We thus obtain 13 (=9+4) galaxies useful for our analysis.

\subsection{Data from K20}
K20 conduct follow up spectroscopy for 8 EMPG candidates with MagE.
All 8 sources are confirmed as star-forming galaxies, and 2 out of the 8 sources are EMPGs with $12+\log(\mathrm{O/H})<7.69$.
These 8 galaxies are useful for outflow studies in a low-mass regime because of their low stellar masses ($\log [M_*/M_\odot] =5-7$) and active star formation. 
The physical properties of these galaxies are taken from K20 and listed in Table \ref{tab:sample}.

Finally, we combine the K20 spectra (8 galaxies) with newly observed data (13 galaxies) and obtain a total of 21 (=8+13) galaxies for our sample as listed in Table \ref{tab:sample}.
In Figure \ref{fig:sfr_m}, we show the SFRs and stellar masses of our galaxies which are derived in Section \ref{galaxy_property}.
We also include the galaxies investigated by several previous outflows studies.
Our galaxies have high specific star formation rate ($\mathrm{sSFR=SFR}/M_*$) suggestive of active star formation that may lead to observable outflow signatures.

\begin{figure*}[ht!]
\includegraphics[width=\textwidth]{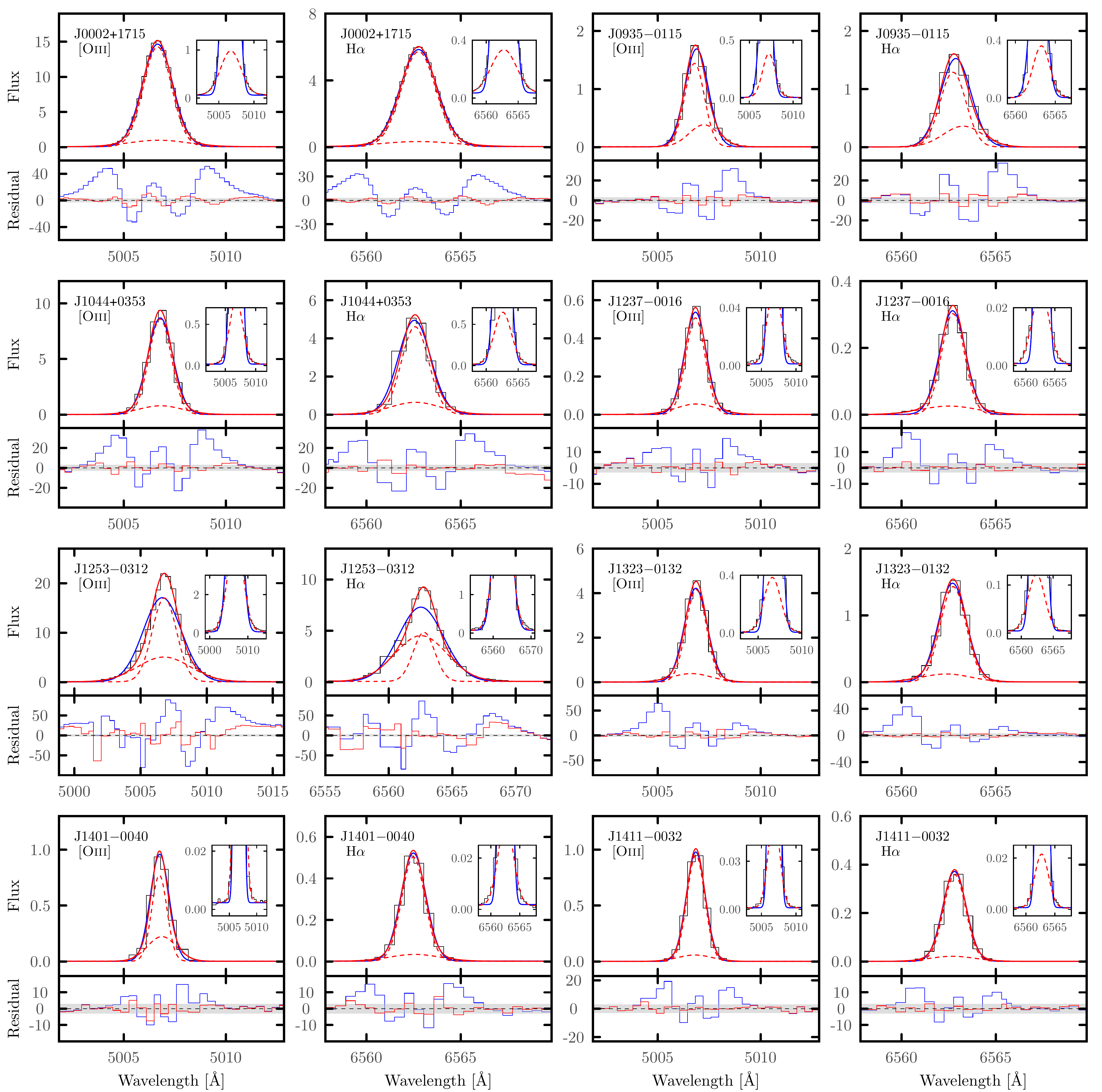}
\caption{Spectra of H$\alpha$ and [{\sc Oiii}] lines with the best-fit profiles.
We only show the spectra for the galaxies that have double-Gaussian profiles in at least one of the H$\alpha$ and [{\sc Oiii}] lines.
Each panel has two sub-panels presenting the emission lines (top) and the fitting residuals (bottom).
In the top panels, the black histograms indicate the observed spectra, while the blue- and red-solid lines are the best-fit double and single Gaussian profiles, respectively.
The red-dashed lines denote the individual narrow and broad components of the double-Gaussian profiles. 
The flux densities are given in units of $10^{-17}\mathrm{erg~s^{-1}~cm^{-2}~\AA^{-1}}$.
The wavelengths are in rest frame.
For the panels with H$\alpha$ lines, the nearby [{\sc Nii}] lines are outside the wavelength range and have no effect on the fitting of the H$\alpha$ lines.
The inset panels show the zoom-in of the spectra, the best-fit single Gaussian profiles, and the broad components.
In the bottom panels, we show fitting residuals normalized by 1$\sigma$ flux errors, with the blue and red histograms for the single and double Gaussian profiles, respectively.
The shade regions indicate the $3\sigma$ errors.}
\label{fig:fitting1}
\end{figure*}

\begin{figure*}[ht]
\includegraphics[width=\textwidth]{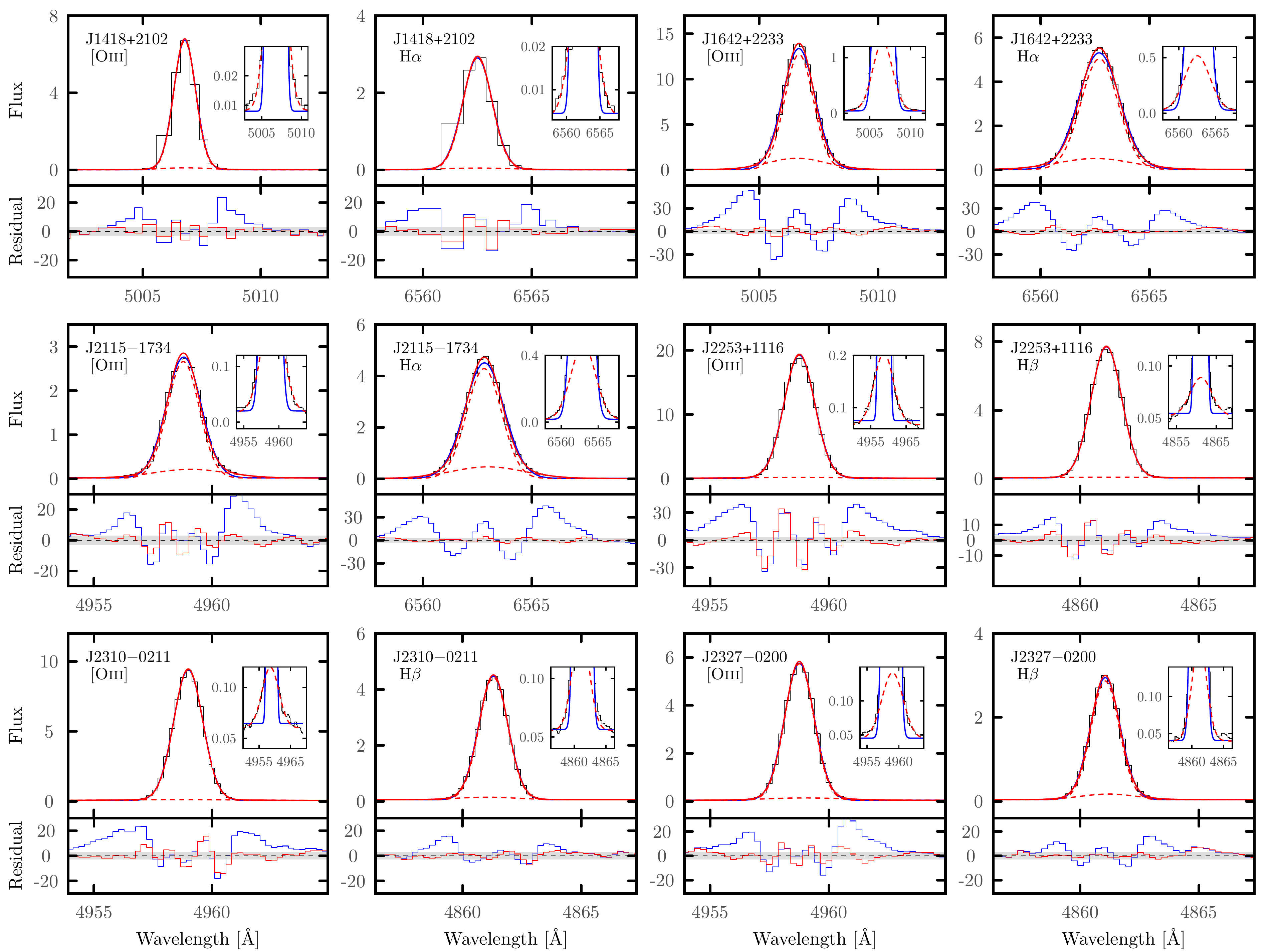}
\caption{Figure \ref{fig:fitting1} continued}
\label{fig:fitting2}
\end{figure*}

\begin{figure}[ht!]
    \centering
    \gridline{
    \fig{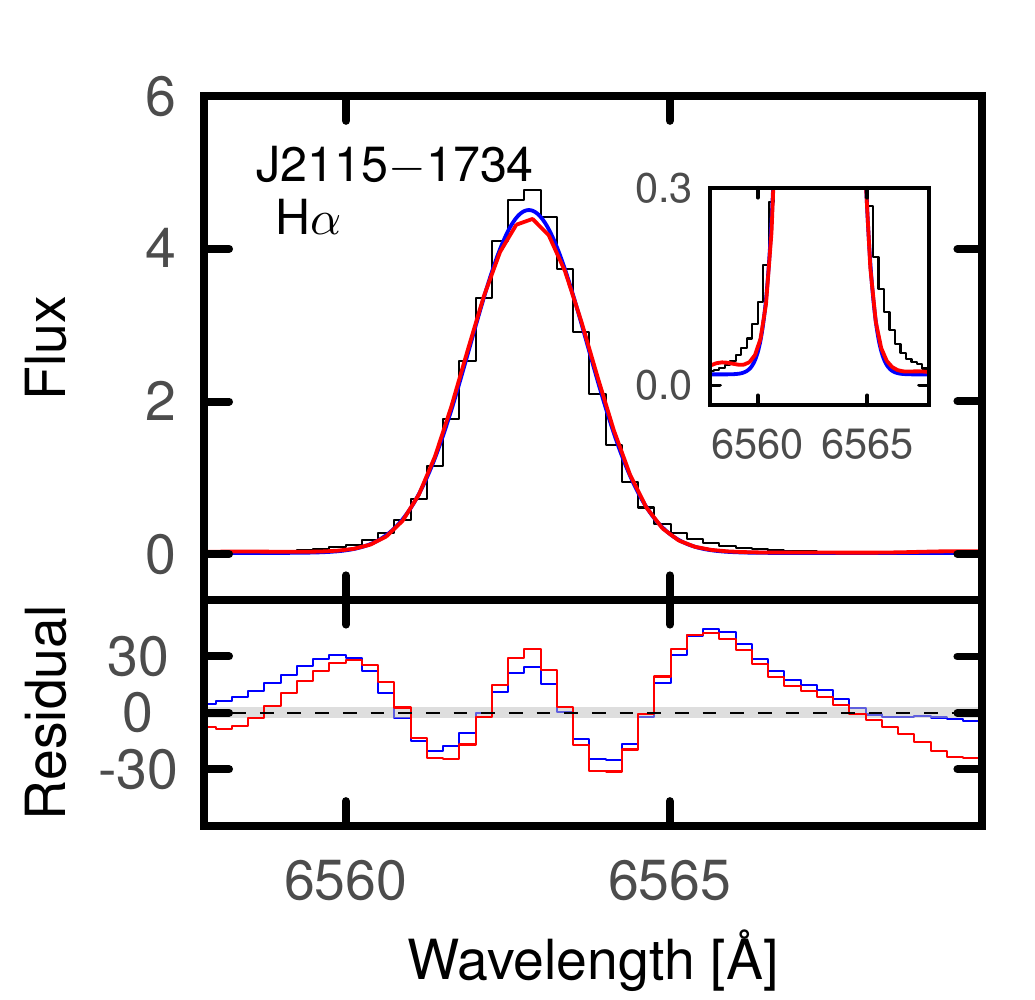}{0.24\textwidth}{}
    \fig{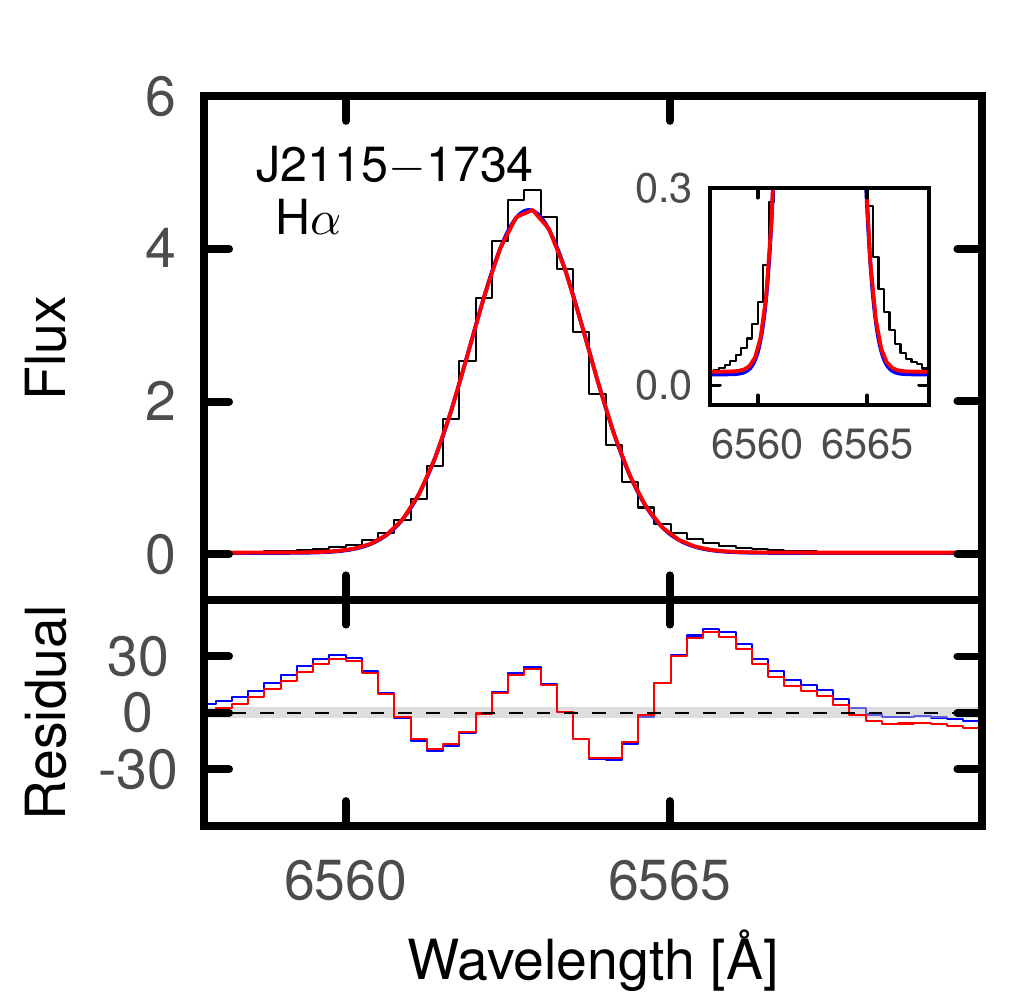}{0.24\textwidth}{}
    }
    \caption{\textit{left}: Best-fit single-Gaussian profile. For the blue line, no instrumental broadening is taken into account, while for the red line, the instrumental profile is convoluted before fitting. \textit{right}: Same as the left panel but now the red line includes the effect of natural broadening instead of instrumental broadening.}
    \label{fig:fitting_justification}
\end{figure}

\section{Analysis}
\label{analysis}
\subsection{Spectrum Fitting}
\label{fitting1d}
We investigate the line profiles of H$\alpha$ and [{\sc Oiii}]$\lambda 5007$ emission lines, both of which have been used to investigate outflows driven by star-formation in previous studies \citep[e.g.,][]{Cicone+16,Bruno+19}.
The K20 galaxies include 4 galaxies with saturated lines of [{\sc Oiii}]$\lambda 5007$ and/or H$\alpha$.
In such cases, assuming that the H$\beta$ ([{\sc Oiii}]$\lambda 4959$) line have similar profiles as the H$\alpha$ ([{\sc Oiii}]$\lambda 5007$) line, we investigate the profiles of the H$\beta$ ([{\sc Oiii}]$\lambda 4959$) lines instead.
We investigate the profiles of H$\alpha$ lines without considering the nearby [{\sc Nii}]$\lambda\lambda6548,6583$ lines because the H$\alpha$ lines are relatively narrow and clearly separated from the [{\sc Nii}] lines.
We fit the emission lines with Gaussian profiles, using a customized non-linear least squares fitting routine with the package \texttt{minpack.lm} in the R language.
For each Gaussian profile, there are four free parameters, the amplitude, line width ($\sigma_\mathrm{obs}$), line central wavelength, and one parameter for the flat continuum where the first three parameters define a Gaussian shape.
We do not fix the line central wavelengths to the systemic redshifts determined by multiple emission lines (Section \ref{galaxy_property}), because some of the emission lines may have shifts from the systemic redshifts. 
By looking at the fitting residuals, we find that the emission line profiles show broad wings and statistically-significant residuals over the best-fit Gaussian profiles for 14 out of the 21 galaxies which are presented in Figure \ref{fig:fitting1} and \ref{fig:fitting2}.
For the rest of 7 galaxies, we do not find significant broad wings.

The SNRs calculated within 3$\sigma_\mathrm{obs}$ around the central wavelengths are as high as $\sim 1000$ (see Figure \ref{fig:BN_SNR}).
Because the high SNRs may reveal complex structures, we need to consider the possibility that the broad wings presented in the emission lines originate from the instrumental broadening of MagE.
We identify several strong skylines and stack the skylines to approximate the instrumental profile, assuming that the instrumental profile does not depend on wavelength.
We follow the same procedure as the single-Gaussian fitting, except that we convolve the Gaussian profile with the instrumental profile before fitting the parameters.
In the left panel of Figure \ref{fig:fitting_justification}, we show one example spectrum that is fitted by the single-Gaussian profile with and without instrumental broadening.
Whether or not we take into account the instrumental broadening, the best-fit single-Gaussian profile shows broad wings and large residuals.
We thus conclude that the broad wings cannot be dominated by the instrumental profile.

Another possible origin of the broad wings is the intrinsic natural broadening of emission lines that has a Lorentzian profile.
We similarly fit the emission lines with the convolution of Gaussian and Lorentzian profiles, which is known as the Voigt profile.
We fix the line width of the Lorentzian profile as the typical value produced by the natural broadening of H$\alpha$ that is $\mathrm{FWHM}\sim4.6\times10^{-3}~\mathrm{\AA}$.
As shown in the right panel of Figure \ref{fig:fitting_justification}, the best-fit single Gaussian profile with natural broadening cannot explain the broad wings.
Although the pressure broadening from collision of atoms also has a Lorentzian profile, it is negligible in nebular gas.

Ruling out the possibilities of instrumental broadening and natural broadening, it is likely that the emission lines consist of multiple Gaussian components \citep[e.g.,][]{Newman+12,Freeman+19}.
% \st{Here we assume that the emission lines consist of narrow and broad lines produced in star-forming regions and outflowing nebulae, respectively.
% We do not consider the possibility that the broad lines originate from AGN accretion disks because the broad wings are found in both the H$\alpha$ lines and the forbidden [{\sc Oiii}] lines.}
We conduct double-Gaussian profile fitting using a total of seven free parameters, the flat continuum and two sets of three Gaussian parameters, namely the amplitude, central wavelength, and line width.
From the double-Gaussian profile fitting, we obtain two Gaussian profiles with different line widths.  
We refer to the narrower (broader) Gaussian profile as a narrow (broad) component.
We allow for a velocity difference ($\Delta v$) between the central wavelengths of the narrow and broad components.
In the spectra shown in the top panels of Figures \ref{fig:fitting1} and \ref{fig:fitting2}, the narrow/broad components and overall shapes of double-Gaussian profiles are overplotted with dashed and solid red lines, respectively.
The double-Gaussian profiles explain well the line shapes, leaving very small residuals. 
We thus do not increase the number of components to three and beyond.
Finally, we conclude that the line shapes of the 14 galaxies are well explained by the double-Gaussian profile.
For the following analysis, we assume that the broad components trace outflowing gas.
In Section \ref{profile_discuss} we discuss alternative interpretations of the double-Gaussian profile and show that the broad components are most likely originated from outflows.

Table \ref{tab:line} presents the best-fit parameters of the double-Gaussian profiles. 
The line widths listed in Table \ref{tab:line} are not corrected for instrumental broadening.
In further analysis, we calculate the intrinsic-line widths by quadratically subtracting $\sigma_\mathrm{inst}$ from the $\sigma_\mathrm{n}$ and $\sigma_\mathrm{b}$ values in Table \ref{tab:line}.
For our MagE spectra taken with the slit widths of $0.''7$, we evaluate $\sigma_\mathrm{inst}$ to be $27.8~\mathrm{km~s^{-1}}$ in the same manner as K20 who use unresolved emission lines of the lamp data. 
For the K20 spectra, we apply $\sigma_\mathrm{inst}$ values obtained by K20 that are $26.4$ and $33.3~\mathrm{km~s^{-1}}$ for the slit widths of $0.''85$ and $1.''20$, respectively.

\subsection{Galaxy Properties}
\label{galaxy_property}

\begin{deluxetable*}{llcCCCCCC}
    % \tablenum{2}
    \tablecaption{Emission Line Fitting Results with Double-Gaussian Profiles}
    % \tablewidth{0pt}
    \tablehead{
        \colhead{ID} & \colhead{line} & \colhead{$\lambda_0$} &
        % \colhead{$\lambda$} & 
        \colhead{$F_\mathrm{n}$} & 
        \colhead{$\sigma_\mathrm{n}$} & \colhead{$\lambda_\mathrm{n}$} &
        \colhead{$F_\mathrm{b}$} & 
        \colhead{$\sigma_\mathrm{b}$} & \colhead{$\lambda_\mathrm{b}$} \\
        \colhead{} & \colhead{} & \colhead{(\AA)} &  
        \colhead{($\mathrm{erg~s^{-1}~cm^{-2}}$)} &
        \colhead{($\mathrm{km~s^{-1}}$)} &
        \colhead{(\AA)} & 
        \colhead{($\mathrm{erg~s^{-1}~cm^{-2}}$)} &
        \colhead{($\mathrm{km~s^{-1}}$)} &
        \colhead{(\AA)}\\
        \colhead{(1)} & \colhead{(2)} & \colhead{(3)} & 
        \colhead{(4)} & \colhead{(5)} & \colhead{(6)} & 
        \colhead{(7)} & \colhead{(8)} & \colhead{(9)}
    }
    % decimalcolnumbers
    \startdata
J0002$+$1715 & H$\alpha$ & 6562.82 & 2.93\pm0.01 & 41.86\pm0.12 & 6699.47 & 0.34\pm0.02 & 90.96\pm1.67 & 6699.46 \\
  & [{\sc Oiii}] & 5006.84 & 5.50\pm0.06 & 41.24\pm0.28 & 5110.94 & 0.74\pm0.09 & 86.29\pm0.28 & 5110.99 \\
J0935$-$0115 & H$\alpha$ & 6562.82 & 2.30\pm0.08 & 31.19\pm0.53 & 6669.12 & 0.99\pm0.12 & 47.92\pm0.86 & 6669.65 \\
  & [{\sc Oiii}] & 5006.84 & 1.80\pm0.11 & 28.63\pm0.76 & 5087.95 & 0.72\pm0.15 & 43.22\pm0.76 & 5088.41 \\
J1044$+$0353 & H$\alpha$ & 6562.82 & 2.77\pm0.09 & 32.98\pm0.61 & 6649.01 & 0.70\pm0.14 & 60.56\pm2.92 & 6649.03 \\
  & [{\sc Oiii}] & 5006.84 & 3.92\pm0.10 & 32.77\pm0.51 & 5072.72 & 0.70\pm0.15 & 64.03\pm0.51 & 5072.75 \\
J1237$-$0016 & H$\alpha$ & 6562.82 & 3.63\pm0.05 & 27.95\pm0.27 & 6893.82 & 0.75\pm0.09 & 68.57\pm3.62 & 6893.60 \\
  & [{\sc Oiii}] & 5006.84 & 4.66\pm0.11 & 28.11\pm0.42 & 5259.42 & 1.05\pm0.20 & 58.73\pm0.42 & 5259.47 \\
J1253$-$0312 & H$\alpha$ & 6562.82 & 1.36\pm0.23 & 38.78\pm4.61 & 6713.78 & 3.07\pm0.40 & 93.93\pm3.25 & 6713.54 \\
  & [{\sc Oiii}] & 5006.84 & 5.22\pm0.32 & 55.06\pm2.19 & 5122.01 & 3.34\pm0.44 & 119.50\pm2.19 & 5121.97 \\
J1323$-$0132 & H$\alpha$ & 6562.82 & 3.31\pm0.04 & 31.06\pm0.21 & 6712.04 & 0.52\pm0.05 & 58.99\pm1.44 & 6711.67 \\
  & [{\sc Oiii}] & 5006.84 & 7.04\pm0.13 & 29.66\pm0.35 & 5120.77 & 1.23\pm0.18 & 56.99\pm0.35 & 5120.54 \\
J1401$-$0040 & H$\alpha$ & 6562.82 & 3.43\pm0.05 & 28.09\pm0.28 & 6639.14 & 0.48\pm0.09 & 60.75\pm3.43 & 6639.22 \\
  & [{\sc Oiii}] & 5006.84 & 3.20\pm0.20 & 22.43\pm0.90 & 5065.22 & 1.74\pm0.33 & 42.49\pm0.90 & 5065.35 \\
J1411$-$0032 & H$\alpha$ & 6562.82 & 3.60\pm0.05 & 26.85\pm0.21 & 6734.56 & 0.43\pm0.09 & 54.18\pm3.03 & 6734.45 \\
  & [{\sc Oiii}] & 5006.84 & 7.02\pm0.11 & 25.78\pm0.24 & 5137.88 & 0.82\pm0.18 & 50.15\pm0.24 & 5137.84 \\
J1418$+$2102 & H$\alpha$ & 6562.82 & 3.48\pm0.04 & 30.37\pm0.22 & 6620.85 & 0.13\pm0.04 & 75.09\pm7.47 & 6620.86 \\
  & [{\sc Oiii}] & 5006.84 & 5.56\pm0.09 & 27.54\pm0.28 & 5051.29 & 0.19\pm0.11 & 63.63\pm0.28 & 5051.34 \\
J1642$+$2233 & H$\alpha$ & 6562.82 & 2.80\pm0.03 & 38.84\pm0.28 & 6675.91 & 0.55\pm0.06 & 77.58\pm2.09 & 6675.71 \\
  & [{\sc Oiii}] & 5006.84 & 5.06\pm0.06 & 36.41\pm0.27 & 5093.04 & 0.98\pm0.10 & 72.15\pm0.27 & 5092.95 \\
J2115$-$1734 & H$\alpha$ & 6562.82 & 2.86\pm0.03 & 36.48\pm0.21 & 6713.49 & 0.63\pm0.04 & 76.07\pm1.45 & 6713.67 \\
  & [{\sc Oiii}] & 4958.91 & 1.49\pm0.02 & 40.66\pm0.38 & 5072.64 & 0.25\pm0.03 & 92.62\pm4.16 & 5072.95 \\
J2253$+$1116 & H$\beta$ & 4861.33 & 1.00\pm0.00 & 39.73\pm0.11 & 4896.59 & 0.02\pm0.00 & 164.66\pm21.18 & 4896.68 \\
  & [{\sc Oiii}] & 4958.91 & 2.50\pm0.02 & 38.96\pm0.16 & 4994.96 & 0.07\pm0.01 & 163.34\pm16.37 & 4994.88 \\
J2310$-$0211 & H$\beta$ & 4861.33 & 0.97\pm0.01 & 37.16\pm0.14 & 4921.82 & 0.04\pm0.01 & 79.19\pm4.99 & 4921.56 \\
  & [{\sc Oiii}] & 4958.91 & 2.11\pm0.01 & 37.07\pm0.09 & 5020.74 & 0.06\pm0.01 & 161.18\pm11.42 & 5020.37 \\
J2327$-$0200 & H$\beta$ & 4861.33 & 0.93\pm0.01 & 36.54\pm0.22 & 4949.13 & 0.09\pm0.02 & 71.96\pm3.96 & 4949.28 \\
  & [{\sc Oiii}] & 4958.91 & 1.93\pm0.01 & 36.83\pm0.13 & 5048.61 & 0.09\pm0.01 & 101.26\pm5.44 & 5048.86
    \enddata
    \tablecomments{Columns: (1) ID. (2) Name of emission line. (3) Rest-frame wavelength in air. \revised{(4)-(6) Line flux normalized to H$\beta$, line width, and observed center wavelength of the narrow components. (7)-(9) same as Columns (4)-(6) but for the broad components.}}
    % \end{tabular}
    \label{tab:line}
\end{deluxetable*}

\begin{deluxetable*}{cccccccc}
    % \tablenum{2}
    \tablecaption{Photometry used for SED fitting}
    \tablehead{
        \colhead{ID} & 
        \colhead{FUV} & \colhead{NUV} &
        \colhead{u} & \colhead{g} &
        \colhead{r} & \colhead{i} &
        \colhead{z} \\
        \nocolhead{} & 
        \colhead{(mag)} & \colhead{(mag)} &
        \colhead{(mag)} & \colhead{(mag)} &
        \colhead{(mag)} & \colhead{(mag)} &
        \colhead{(mag)} \\
        \nocolhead{(1)} & 
        \colhead{(2)} & \colhead{(3)} &
        \colhead{(4)} & \colhead{(5)} &
        \colhead{(6)} & \colhead{(7)} &
        \colhead{(8)}
    }
    % \decimalcolnumbers
    \startdata
    %\multicolumn{4}{c} &
& & & & \multicolumn{4}{c}{HSC photometry}\\
\cline{5-8}
J0845$+$0131 & 21.64 $\pm$ 0.11 & 21.60 $\pm$ 0.10 & 22.16 $\pm$ 0.29 & 21.12 $\pm$ 0.00 & 21.19 $\pm$ 0.00 & 21.97 $\pm$ 0.01 & 21.99 $\pm$ 0.01 \\
J0912$-$0104 & 22.32 $\pm$ 0.19 & 22.93 $\pm$ 0.34 & 22.31 $\pm$ 0.32 & 22.22 $\pm$ 0.01 & 22.30 $\pm$ 0.01 & 22.93 $\pm$ 0.01 & 23.02 $\pm$ 0.04 \\
J0935$-$0115 & 19.28 $\pm$ 0.03 & 19.26 $\pm$ 0.02 & 19.39 $\pm$ 0.03 & 18.96 $\pm$ 0.00 & 19.05 $\pm$ 0.00 & 19.66 $\pm$ 0.00 & 19.70 $\pm$ 0.00 \\
J1210$-$0103 & 22.93 $\pm$ 0.30 & 22.73 $\pm$ 0.25 & \nodata & 22.92 $\pm$ 0.01 & 22.77 $\pm$ 0.02 & 22.88 $\pm$ 0.02 & 22.86 $\pm$ 0.04 \\
J1237$-$0016 & 21.17 $\pm$ 0.06 & 21.55 $\pm$ 0.05 & 21.68 $\pm$ 0.11 & 20.63 $\pm$ 0.00 & 21.30 $\pm$ 0.00 & 22.15 $\pm$ 0.01 & 22.15 $\pm$ 0.01 \\
J1401$-$0040 & 19.63 $\pm$ 0.04 & 19.74 $\pm$ 0.02 & 20.40 $\pm$ 0.04 & 19.65 $\pm$ 0.00 & 19.91 $\pm$ 0.00 & 20.52 $\pm$ 0.00 & 20.41 $\pm$ 0.00 \\
J1407$-$0047 & \nodata & \nodata & 22.61 $\pm$ 0.26 & 22.02 $\pm$ 0.01 & 22.16 $\pm$ 0.01 & 22.72 $\pm$ 0.01 & 22.67 $\pm$ 0.02 \\
J1411$-$0032 & \nodata & \nodata & 21.44 $\pm$ 0.19 & 20.50 $\pm$ 0.00 & 21.39 $\pm$ 0.00 & 22.24 $\pm$ 0.01 & 22.13 $\pm$ 0.01 \\
J1452$+$0241 & 20.63 $\pm$ 0.05 & 20.82 $\pm$ 0.06 & 21.30 $\pm$ 0.13 & 20.74 $\pm$ 0.00 & 20.78 $\pm$ 0.00 & 21.70 $\pm$ 0.01 & 21.79 $\pm$ 0.01 \\
& & & & \multicolumn{4}{c}{SDSS photometry}\\
\cline{5-8}
J1044$+$0353 & 18.00 $\pm$ 0.06 & 17.82 $\pm$ 0.02 & 18.27 $\pm$ 0.02 & 17.40 $\pm$ 0.00 & 17.65 $\pm$ 0.01 & 18.66 $\pm$ 0.01 & 18.74 $\pm$ 0.04 \\
J1253$-$0312 & 16.27 $\pm$ 0.03 & 16.17 $\pm$ 0.02 & 15.93 $\pm$ 0.00 & 15.06 $\pm$ 0.00 & 15.57 $\pm$ 0.00 & 16.06 $\pm$ 0.00 & 15.97 $\pm$ 0.01 \\
J1323$-$0132 & 18.76 $\pm$ 0.10 & 18.53 $\pm$ 0.04 & 19.09 $\pm$ 0.02 & 18.06 $\pm$ 0.01 & 18.78 $\pm$ 0.01 & 19.44 $\pm$ 0.03 & 19.28 $\pm$ 0.09 \\
J1418$+$2102 & \nodata & \nodata & 18.20 $\pm$ 0.02 & 17.51 $\pm$ 0.01 & 17.58 $\pm$ 0.01 & 17.91 $\pm$ 0.01 & 17.62 $\pm$ 0.04 \\
    \enddata
    \tablecomments{Columns: (1) ID. (2) FUV magnitude measured with the GALEX data. (3) Same as (2), but for NUV magnitude. (4) SDSS $u$-band magnitude. \revised{In case the galaxy only has HSC photometry, the value is calculated from the $u-g$ color of SDSS and the $g$-band magnitude of HSC. See text} (5)-(8) $griz$ magnitudes of HSC or SDSS photometry. \revised{The symbols `\nodata' indicate the lack of photometry data.}}
    % \tablecomments{xxxx}
    % \end{tabular}
    \label{tab:photometry}
\end{deluxetable*}

\begin{deluxetable*}{clrrrrrrr}
    % \tablenum{2}
    \tablecaption{Outflow Properties}
    \tablewidth{0pt}
    \tablehead{
        \colhead{ID} & \colhead{line} &
        \colhead{$BNR$} & \colhead{$\Delta v$} &
        \colhead{FWHM$_\mathrm{b}$} & \colhead{$v_\mathrm{max}$} & 
        \colhead{$v_\mathrm{esc}$} & 
        \colhead{$\eta$} & \colhead{$\eta_\mathrm{max}$} \\
        \nocolhead{} & \nocolhead{} & \nocolhead{} &
        \colhead{($\mathrm{km~s^{-1}}$)} &
        \colhead{($\mathrm{km~s^{-1}}$)} & \colhead{($\mathrm{km~s^{-1}}$)} &
        \nocolhead{} & \nocolhead{} & \nocolhead{} \\
        \colhead{(1)} & \colhead{(2)} & \colhead{(3)} &
        \colhead{(4)} & \colhead{(5)} &
        \colhead{(6)} & \colhead{(7)} &
        \colhead{(8)} & \colhead{(9)}
    }
    % \decimalcolnumbers
    \startdata
J0002$+$1715 & H$\alpha$    & 0.12 $\pm$ 0.01 & -0.28  $\pm$ 0.82  & 199.32 $\pm$ 3.65  & 99.94  $\pm$ 2.00  & $86.03_{-14.85}^{+17.95}$ & 0.65 & 16.93 \\
             & [{\sc Oiii}] & 0.14 $\pm$ 0.02 & 3.06   $\pm$ 1.55  & 187.46 $\pm$ 6.28  & 96.79  $\pm$ 3.50  &        &      \\
J0935$-$0115 & H$\alpha$    & 0.43 $\pm$ 0.05 & 24.27  $\pm$ 2.37  & 91.91  $\pm$ 1.65  & 70.23  $\pm$ 2.51  & $68.86_{-13.24}^{+17.16}$ & 1.05 & 34.24\\
             & [{\sc Oiii}] & 0.40 $\pm$ 0.09 & 27.17  $\pm$ 5.17  & 77.94  $\pm$ 2.55  & 66.14  $\pm$ 5.32  &        &      \\
J1044$+$0353 & H$\alpha$    & 0.25 $\pm$ 0.05 & 0.89   $\pm$ 1.11  & 126.7  $\pm$ 6.11  & 64.24  $\pm$ 3.25  & $66.82_{-12.01}^{+14.67}$ & 0.44 & 21.04 \\
             & [{\sc Oiii}] & 0.18 $\pm$ 0.04 & 1.49   $\pm$ 1.73  & 135.82 $\pm$ 6.99  & 69.40  $\pm$ 3.90  &        &      \\
J1237$-$0016 & H$\alpha$    & 0.21 $\pm$ 0.02 & -9.26  $\pm$ 1.70  & 147.60 $\pm$ 7.79  & 83.06  $\pm$ 4.25  & $70.02_{-12.07}^{+14.63}$ & 0.71 & 23.12 \\
             & [{\sc Oiii}] & 0.23 $\pm$ 0.04 & 2.45   $\pm$ 1.72  & 121.83 $\pm$ 6.62  & 63.36  $\pm$ 3.73  &        &      \\
J1253$-$0312 & H$\alpha$    & 2.26 $\pm$ 0.49 & -10.83 $\pm$ 4.24  & 211.29 $\pm$ 7.31  & 116.47 $\pm$ 5.60  & $94.56_{-16.64}^{+20.17}$ & 1.40 & 131.39 \\
             & [{\sc Oiii}] & 0.64 $\pm$ 0.09 & -1.88  $\pm$ 3.16  & 273.68 $\pm$ 9.35  & 138.72 $\pm$ 5.64  &        &      \\
J1323$-$0132 & H$\alpha$    & 0.16 $\pm$ 0.02 & -16.53 $\pm$ 1.39  & 122.52 $\pm$ 2.98  & 77.79  $\pm$ 2.04  & $69.38_{-12.21}^{+14.96}$ & 0.21 & 17.25 \\
             & [{\sc Oiii}] & 0.18 $\pm$ 0.03 & -13.8  $\pm$ 1.69  & 117.15 $\pm$ 3.96  & 72.38  $\pm$ 2.60  &        &      \\
J1401$-$0040 & H$\alpha$    & 0.14 $\pm$ 0.03 & 3.64   $\pm$ 1.86  & 127.19 $\pm$ 7.18  & 67.23  $\pm$ 4.04  & $58.43_{-10.81}^{+13.15}$ & 0.41 & 13.50 \\
             & [{\sc Oiii}] & 0.54 $\pm$ 0.11 & 7.6    $\pm$ 1.95  & 75.68  $\pm$ 3.35  & 45.44  $\pm$ 2.57  &        &      \\
J1411$-$0032 & H$\alpha$    & 0.12 $\pm$ 0.02 & -5.09  $\pm$ 1.67  & 109.5  $\pm$ 6.12  & 59.84  $\pm$ 3.48  & $62.63_{-11.11}^{+13.64}$ & 0.32 & 10.32 \\
             & [{\sc Oiii}] & 0.12 $\pm$ 0.03 & -2.54  $\pm$ 1.48  & 98.28  $\pm$ 5.64  & 51.68  $\pm$ 3.18  &        &      \\
J1418$+$2102 & H$\alpha$    & 0.04 $\pm$ 0.01 & 0.70   $\pm$ 4.16  & 164.26 $\pm$ 16.33 & 82.83  $\pm$ 9.16  & $69.61_{-13.73}^{+19.30}$ & 0.36 & 4.78 \\
             & [{\sc Oiii}] & 0.03 $\pm$ 0.02 & 2.98   $\pm$ 6.10  & 134.79 $\pm$ 21.03 & 70.37  $\pm$ 12.16 &        &      \\
J1642$+$2233 & H$\alpha$    & 0.20 $\pm$ 0.02 & -8.91  $\pm$ 1.23  & 171.78 $\pm$ 4.63  & 94.8   $\pm$ 2.62  & $67.16_{-12.95}^{+14.02}$ & 2.18 & 25.31 \\
             & [{\sc Oiii}] & 0.19 $\pm$ 0.02 & -5.14  $\pm$ 1.09  & 158.13 $\pm$ 4.40  & 84.2   $\pm$ 2.45  &        &      \\
J2115$-$1734 & H$\alpha$    & 0.22 $\pm$ 0.02 & 7.94   $\pm$ 0.88  & 167.99 $\pm$ 3.21  & 91.93  $\pm$ 1.83  & $76.01_{-12.97}^{+15.64}$ & 1.50 & 26.95 \\
             & [{\sc Oiii}] & 0.17 $\pm$ 0.02 & 18.21  $\pm$ 2.95  & 209.05 $\pm$ 9.40  & 122.73 $\pm$ 5.55  &        &      \\
J2253$+$1116 & H$\alpha$    & 0.02 $\pm$ 0.00 & 5.36   $\pm$ 11.44 & 379.73 $\pm$ 48.83 & 195.23 $\pm$ 26.96 & $62.66_{-10.56}^{+12.70}$ & 0.24 & 6.12 \\
             & [{\sc Oiii}] & 0.03 $\pm$ 0.01 & -4.35  $\pm$ 9.15  & 376.57 $\pm$ 37.74 & 192.63 $\pm$ 20.97 &        &      \\
J2310$-$0211 & H$\alpha$    & 0.04 $\pm$ 0.01 & -16.19 $\pm$ 3.57  & 169.19 $\pm$ 10.67 & 100.78 $\pm$ 6.42  & $84.55_{-14.60}^{+17.64}$ & 0.23 & 6.96 \\
             & [{\sc Oiii}] & 0.03 $\pm$ 0.00 & -21.66 $\pm$ 6.79  & 371.36 $\pm$ 26.32 & 207.34 $\pm$ 14.81 &        &      \\
J2327$-$0200 & H$\alpha$    & 0.09 $\pm$ 0.02 & 8.85   $\pm$ 2.20  & 150.21 $\pm$ 8.26  & 83.95  $\pm$ 4.68  & $75.08_{-12.96}^{+15.44}$ & 0.33 & 11.53 \\
             & [{\sc Oiii}] & 0.05 $\pm$ 0.01 & 14.93  $\pm$ 3.47  & 225.18 $\pm$ 12.1  & 127.52 $\pm$ 6.97  &        &        
    \enddata
    \tablecomments{Columns: (1) ID. (2) emission line used to trace outflows. (3) narrow-to-broad flux ratio. (4) velocity shift of broad component respect to the narrow component. (5) FWHM of the broad component. (6) maximum outflow velocity estimated with Equation \ref{eq:vmax}. (7) escape velocity of the galaxy. (8) fiducial mass-loading factor. (9) maximum mass-load factor with extreme parameters.}
    \label{tab:outflow}
\end{deluxetable*}

We estimate the redshifts, dust extinction, stellar masses, stellar ages, and star-formation rates (SFRs) of the thirteen galaxies observed in our MagE run (Table \ref{tab:obsobj}).
For galaxies taken from K20, we adopt these values estimated in K20.
We also estimate the circular velocities of all our galaxies for further analysis.
Discussions of galaxy properties will be presented in the forthcoming paper (K. Nakajima et al. in preparation).

We estimate redshifts, $z$, and color excesses, $E(B-V)$, in the same manner as K20.
We compare the observed central wavelengths of 4 strong emission lines (H$\alpha$, H$\beta$, [{\sc Oiii}]$\lambda 5007$, and [{\sc Oiii}]$\lambda 4959$) with the respective rest-frame wavelengths in air.
Color excesses are calculated from the Balmer decrements of H$\alpha$, H$\beta$, H$\gamma$, H$\delta$, and H$\varepsilon$ under the assumptions of the case B recombination and the dust attenuation curve of the Small Magellanic Cloud \citep{SMC1,SMC2}.

In order to obtain the stellar masses and ages, we use the BayEsian Analysis of GaLaxy sEds (BEAGLE, v0.23.0; \citeauthor{Chevallard+16} \citeyear{Chevallard+16}) to fit photometry data acquired from Galaxy Evolution Explorer (GALEX; \citealt{GALEX}), SDSS, and HSC (Table \ref{tab:photometry}).
For the HSC photometry of $griz$, we use the HSC-SSP internal data of the latest S20A data release, and adopt ``cmodel" photometry.
For the SDSS data ($ugriz$), we use the ``ModelMag" photometry from the DR16 catalog \citep{SDSS16}.
The $u$-band photometry of SDSS is also used for the HSC sources if available, where we assume the ($u-g$) color of SDSS and the $g$-band magnitude of HSC.
For the GALEX photometry of FUV and NUV bands, we choose the deepest images available for each of the sources from the MAST archive \footnote{\url{https://archive.stsci.edu/}}, and obtain the total magnitudes using ``MAG\_AUTO" of SExtractor. 
We do not use the GALEX photometry data for the three galaxies in our sample that are highly blended with nearby sources in the GALEX images.
All magnitudes are corrected for Galactic extinction based on the \cite{milkyway_extinction}'s map as well as the extinction curve of \cite{Cardelli+89}.
The results are listed in Table \ref{tab:photometry}.
Because $g$- and $r$-band magnitudes include contribution from strong emission lines, we choose the spectral energy distribution (SED) templates of \cite{Gutkin+16} that include nebular emission lines.
We adopt the \cite{Chabrier+03} stellar initial mass function (IMF) with upper mass cutoffs $100~M_\odot$.
For the fitting process, we assume a constant star-formation history and a dust extinction curve of \cite{Charlot+00}.
We fit 5 free parameters with uniform prior distributions, the age of star-formation period ($4 < \log[t/\mathrm{yr}] < 10$), stellar mass ($4 < \log[M/M_\odot] < 9$), metallicity ($-2 < \log[\mathrm{Z/Z_\odot}] < 0$), nebular ionization parameter ($-3.0 < \log\mathrm{U} < -0.5$), and $V$-band attenuation optical depth ($0 < \hat{\tau}_V < 3$).
The redshifts are fixed to the values derived from the spectra, that are used to derive the distances of the galaxies during the fitting process.
After we run the SED fitting, we take into account the peculiar motions of galaxies that affect the estimation of distances and stellar masses.
As our galaxies have low redshifts, we consider that the luminosity distances are proportional to $cz-v_\mathrm{pec}$, where $c$ is the speed of light and $v_\mathrm{pec}$ is the line-of-sight velocity of peculiar motion.
For each galaxy, we perform 1000 Monte-Carlo simulations of $v_\mathrm{pec}$ in which the $1\sigma$ error is $\sim300~\mathrm{km~s^{-1}}$ \citep[e.g.,][]{Kessler+09}.
We then apply the correction factors of $(cz-v_\mathrm{pec})^2/(cz)^2$ to the posterior distribution of the stellar mass given by the SED fitting.
We combine the distributions obtained in the 1000 simulations and take the 16th, 50th, and 84th percentiles.
Finally, we obtain estimates of stellar masses and stellar ages in the range of $\sim10^{4.2}-10^{7.4}~M_\odot$ and $\sim10^{6}-10^{8.4}~\mathrm{yr}$, respectively.

We estimate SFRs with the relation determined by \cite{Kennicutt+98}, which assumes a Salpeter IMF:
\begin{equation}
    \mathrm{SFR}~[M_\mathrm{\odot}~\mathrm{yr}^{-1}] = 7.9\times10^{-42} L(H\alpha)~ [\mathrm{erg~s^{-1}}],
    \label{eq:sfr}
\end{equation}
where $L(H\alpha)$ is the intrinsic luminosity of H$\alpha$ emission. 
To calculate the intrinsic luminosities, we use the line fluxes that are estimated from the best-fit Gaussian profiles in Section \ref{fitting1d}.
The distances are estimated from redshifts assuming a flat $\Lambda$CDM cosmology.
We apply aperture correction and dust extinction correction to the line fluxes.
For aperture correction, we first extract $r$-band photometris from the 1D spectra using the python package \texttt{speclite}.
The differences between the extracted photometries and those we list in Table \ref{tab:photometry} are then used as the factors of aperture correction.
For the uncertainties of SFR, we combine the uncertainties of the flux measurements and those originated from galaxy peculiar motion, in a similar manner to the case of $M_*$.
Finally, we divide the SFRs by 1.8 based on the Chabrier IMF \citep[][]{Hsyu+18}.
We obtain $\log(\mathrm{SFR}/M_\odot~\mathrm{yr}^{-1})\sim-3.79-0.31$ for our sample.
The SFRs estimated from the SED fitting tend to be larger than those estimated from $L(H\alpha)$ by $\sim0.1-0.3$ dex.
In particular, the SED fitting overestimate the SFR of J1411$-$0032 by $\sim3$ dex.
% There exist differences of $\sim 0.5$ dex between SFRs estimated from BEAGLE and the dust-corrected H$\alpha$ fluxes.
Because SFRs calculated with BEAGLE can be affected by various fitting parameters including the star-formation timescale, we adopt SFRs estimated from the dust-corrected H$\alpha$ fluxes.

To estimate $v_\mathrm{cir}$, we first convert the stellar mass to the mass of dark matter (DM) halo ($M_\mathrm{h}$) with the $M_*-M_\mathrm{h}$ relation for low-mass galaxies proposed by \cite{Brook+14}:
\begin{equation}
    M_*=\left(\frac{M_\mathrm{h}}{79.6\times10^6~\mathrm{M}_\odot}\right)^{3.1},
    \label{eq:Mh}
\end{equation}
where $M_\mathrm{h}$ is the DM halo mass defined by an overdensity of $\Delta_c=200$.
Then we adopt the equations in \cite{Mo&White+02} for $v_\mathrm{cir}$:
\begin{equation}
    v_\mathrm{cir}=\left(\frac{GM_\mathrm{h}}{r_\mathrm{h}}\right)^{1/2},
\end{equation}
\begin{equation}
    r_\mathrm{h}=\left(\frac{GM_\mathrm{h}}{100\Omega_\mathrm{m}H_0^2}\right)^{1/3}(1+z)^{-1},
\end{equation}
where $r_\mathrm{h}$ is the halo radius.
\cite{Prole+19} derive $M_\mathrm{h}$ for galaxies with $M_*\sim10^5-10^8~M_\odot$ and find good agreement with the $M_*-M_\mathrm{h}$ relation of Equation (\ref{eq:Mh}).
To derive the uncertainties, we adopt the typical scatter of $\Delta\log(M_\mathrm{h}/M_\odot)\sim0.24$ calculated by \cite{Prole+19} for the galaxies with $M_*\sim10^6~M_\odot$.
The uncertainties of $M_*$ and the typical scatter are propagated into the $M_\mathrm{h}$ and $v_\mathrm{cir}$ estimations.

Galaxy properties are summarized in Table \ref{tab:sample}.

\begin{figure*}[!htb]
    \centering
    \includegraphics[width=\textwidth]{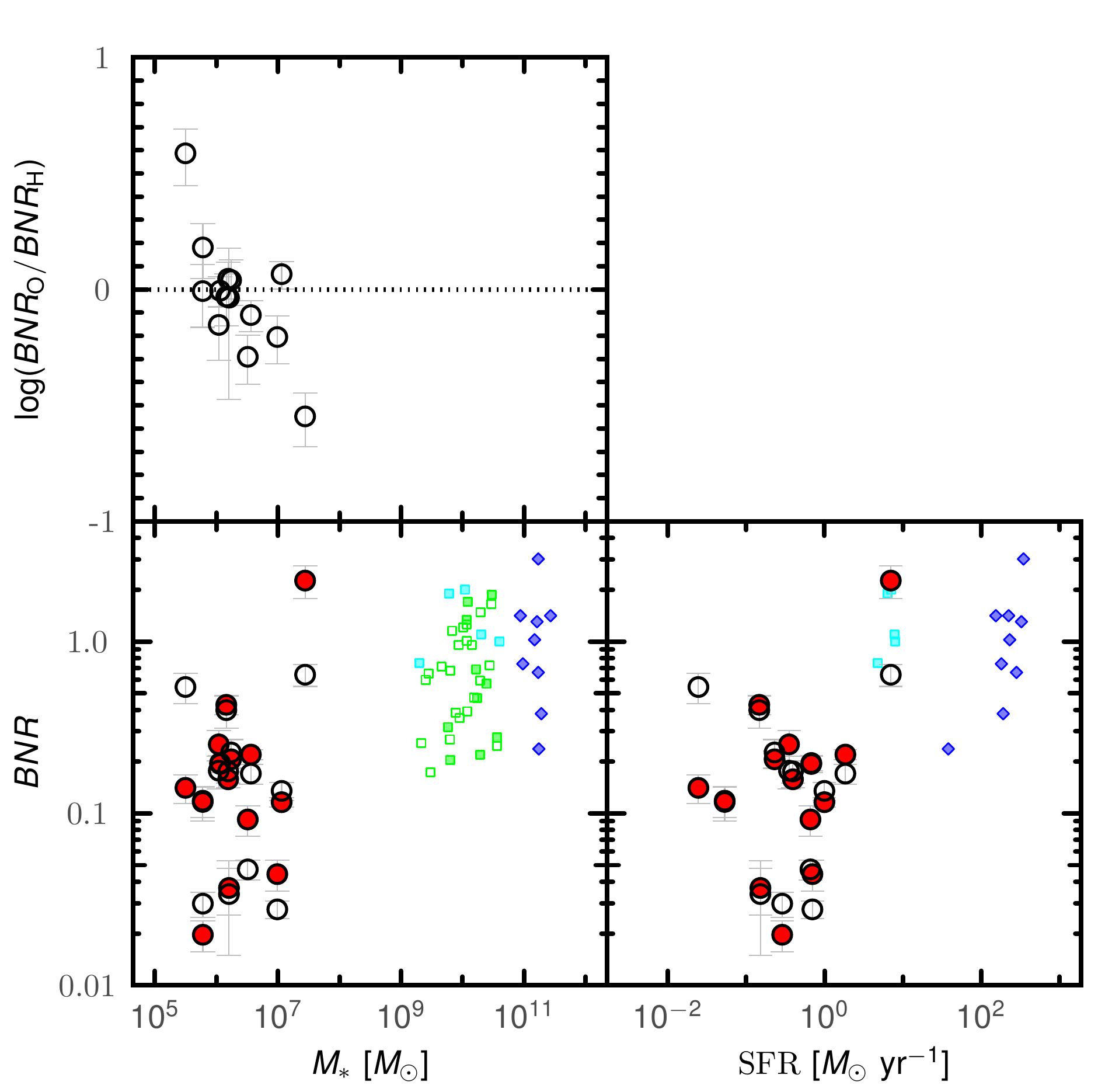}
    \caption{Broad-to-narrow flux ratio ($BNR$). \textit{Bottom left}: $BNR$ as a function of $M_*$. The red filled and open circles are $BNR$ measured with the H$\alpha$ and [{\sc Oiii}] lines, respectively. 
    The green-filled and open squares are taken from \cite{Freeman+19} who evaluate $BNR$ with the H$\alpha$ and [{\sc Oiii}] lines.
    The blue diamonds are calculated from the H$\alpha$ lines by \cite{Perrotta+21}.
    The cyan squares are calculated by \cite{Swinbank+19} with stacked H$\alpha$ spectra.
    \textit{Bottom right}: Same as the bottom left panel, but for SFR. \textit{Top left}: Ratio between the $BNR$ from [{\sc Oiii}] and H$\alpha$ in log scale.}
    \label{fig:BN_mass_sfr}
\end{figure*}

\begin{figure}[!htb]
    \centering
    \includegraphics[width=\linewidth]{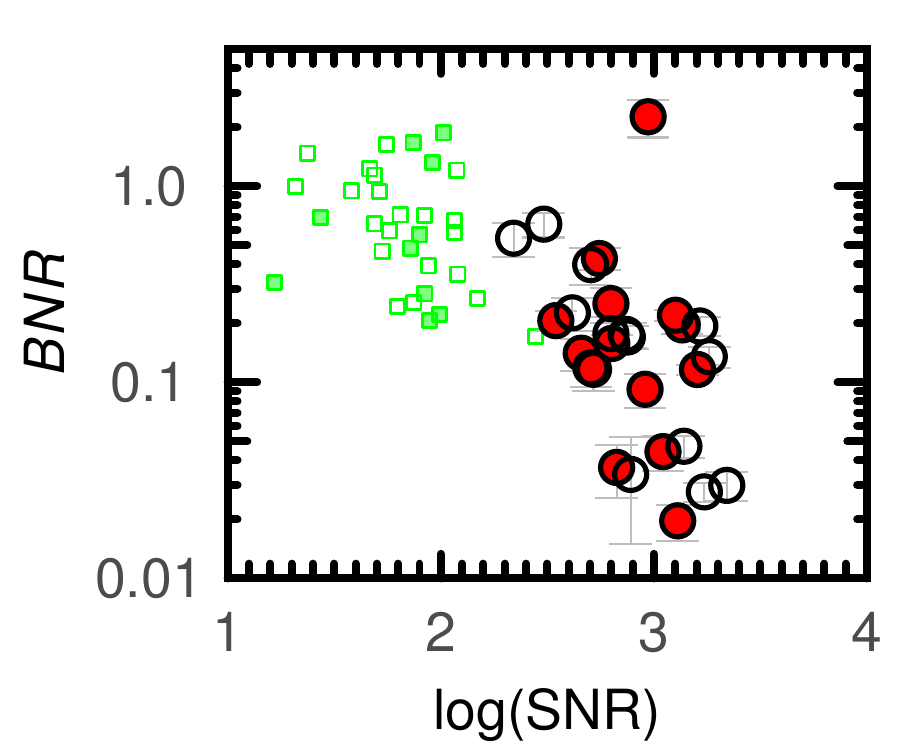}
    \caption{BNR as a function of the SNR of the emission lines. The symbols are the same as Figure \ref{fig:BN_mass_sfr}.}
    \label{fig:BN_SNR}
\end{figure}

\begin{figure*}[!htb]
    \centering
    \includegraphics[width=\textwidth]{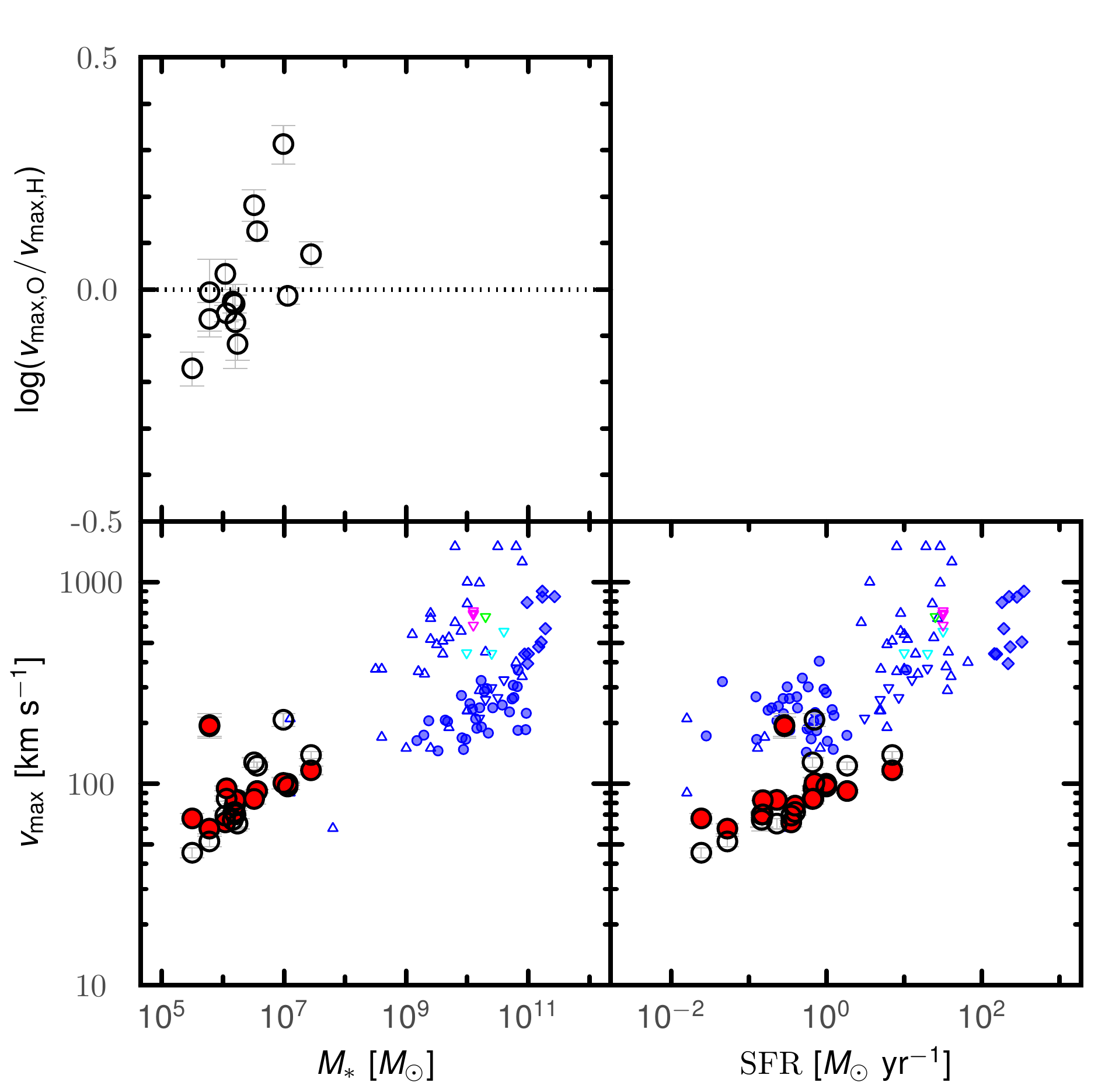}
    \caption{
    Same as Figure \ref{fig:BN_mass_sfr}, but for maximum outflow velocity ($v_\mathrm{max}$).
    The blue circles are taken from \cite{Bruno+19}.
    The other symbols are the same as Figure \ref{fig:sfr_m}.}
    \label{fig:vmax_mass_sfr}
\end{figure*}

\section{Results}
\label{results}
In this study, we investigate ionized outflows with emission lines of [{\sc Oiii}] and H$\alpha$, following the procedures taken by previous studies \citep[e.g.,][]{Concas+17,Freeman+19,Bruno+19}.
Hereafter we focus on the 14 out of 21 galaxies that have emission lines with double-Gaussian profiles.
We characterize outflows with our double-Gaussian profile fitting results, focusing on three properties:
1) the flux ratio between the broad and narrow components ($BNR$), 
2) the velocity shift of the broad component relative to the narrow component ($\Delta v$), 
and 3) the maximum outflow velocity ($v_\mathrm{max}$).
% We evaluate $v_\mathrm{max}$ with the following equation:
We adopt the line-of-sight velocity in the direction of $\Delta v$ as the maximum outflow velocity \citep[][]{Veilleux05}:
\begin{equation}
    v_\mathrm{max}=|\Delta v| + \mathrm{FWHM_b}/2,
    \label{eq:vmax}
\end{equation}
where $\mathrm{FWHM_b}$ is the intrinsic FWHM of the broad component. 
The values of $\mathrm{FWHM_b}$ are calculated from the observed line widths as discussed in Section \ref{fitting1d}.
The outflow properties are listed in Table \ref{tab:outflow}.

We find that 3 (5) out of 14 galaxies have $\Delta v>0$ ($\Delta v<0$) values at the $3\sigma$ levels, indicating that the broad components are redshifted (blueshifted) from the narrow components. 
The other 6 galaxies have $\Delta v$ consistent with zero within the $3\sigma$ errors.
If one assumes symmetric outflows in the line-of-sight direction, the broad component is usually blueshifted due to a strong dust attenuation in the red wing of emission line.
The redshifted broad components with $\Delta v>0$ would therefore suggest that the respective galaxies may have an excess of outflowing gas towards positive redshift.

We obtain $BNR$ values from the H$\alpha$ and [{\sc Oiii}] lines that are referred to as $BNR_\mathrm{H}$ and $BNR_\mathrm{O}$, respectively. 
The top panel of Figure \ref{fig:BN_mass_sfr} shows the ratio of $BNR_\mathrm{H}/BNR_\mathrm{O}$.
We find no significant differences between $BNR_\mathrm{O}$ and $BNR_\mathrm{H}$ due to the moderately large errors of $BNR$.
The results suggest that the broad components in the forbidden [{\sc Oiii}] lines are as prominent as those in the H$\alpha$ lines.
Because the [{\sc Oiii}] emission cannot take place in dense environment, the broad components are unlikely originated from the rotation disks around massive objects, e.g., supermassive black holes (BHs, see Section \ref{profile_discuss} for further discussions).
The bottom panels of Figure \ref{fig:BN_mass_sfr} presents $BNR_\mathrm{H}$ with filled-red circles and $BNR_\mathrm{O}$ with open-black circles.
For both H$\alpha$ and [{\sc Oiii}] emission, we find large scatters of $BNR$ from $\sim0.01$ to $\sim1$.
Similar to the results of previous studies,
we find no clear correlation between $BNR$ and stellar mass (SFR) for our galaxies,
as shown in the bottom left (right) panel of Figure \ref{fig:BN_mass_sfr}.
However, the mean values of $BNR_\mathrm{H}$ and $BNR_\mathrm{O}$ (0.31 and 0.21, respectively) are smaller than the values calculated by \cite{Freeman+19}, 0.76 and 0.74.
The difference of redshifts may not be the reason of different $BNR$ values, because we find no difference of $BNR$ between $z\sim0$ galaxies (\citealt{Perrotta+21}, blue diamonds) and $z\sim2$ galaxies (\citealt{Freeman+19}, green squares).
It is possible that the weak broad components in our galaxies originate from weak outflows with small mass outflow rate.
We will come back to this in Section \ref{eta}.
Another factor we need to consider is the close relation between BNR and the SNR of emission lines.
We show BNR as a function of the emission line SNRs in Figure \ref{fig:BN_SNR}.
The SNRs of our data are significantly higher than those of \cite{Freeman+19}, which allows us to detect low BNR values.

We also investigate the outflow velocities by obtaining $v_\mathrm{max}$ values from the H$\alpha$ and [{\sc Oiii}] lines that are referred to as $v_\mathrm{max,H}$ and $v_\mathrm{max,O}$, respectively. 
In the bottom panels of Figure \ref{fig:vmax_mass_sfr}, we present the results and find positive dependence of $v_\mathrm{max}$ on stellar masses and SFRs. 
The majority of our galaxies have $v_\mathrm{max,H}\sim60-120~\mathrm{km~s^{-1}}$ and $v_\mathrm{max,O}\sim60-140~\mathrm{km~s^{-1}}$.
J2253$+$1116 has relatively fast outflows with $v_\mathrm{max}\sim200~\mathrm{km~s^{-1}}$ in both H$\alpha$ and [{\sc Oiii}] lines. 
Another interesting galaxy is J2310$-$0211 that has $v_\mathrm{max}\sim200~\mathrm{km~s^{-1}}$ only in the [{\sc Oiii}] line.

Similar to $BNR$, we investigate the difference of the H$\alpha$ and [{\sc Oiii}] lines in tracing outflow velocity, as shown in the top left panel of Figure \ref{fig:vmax_mass_sfr}.
% In the top left panel of Figure \ref{fig:vmax_mass_sfr}, 
Four galaxies show the $v_\mathrm{max}$ values of the [{\sc Oiii}] lines to be larger than those of the H$\alpha$ lines.
In particular, J2253$+$1116 has $v_\mathrm{max,O}\sim2v_\mathrm{max,H}$.
Previous studies also report differences between $v_\mathrm{max,O}$ and $v_\mathrm{max,H}$.
\cite{Cicone+16} find outflow velocities inferred from
the [{\sc Oiii}] lines higher than those from the H$\alpha$ lines for local star forming galaxies.

\section{Discussion}
\label{discussion}

\begin{figure}[t!]
    \centering
    \includegraphics[width=\linewidth]{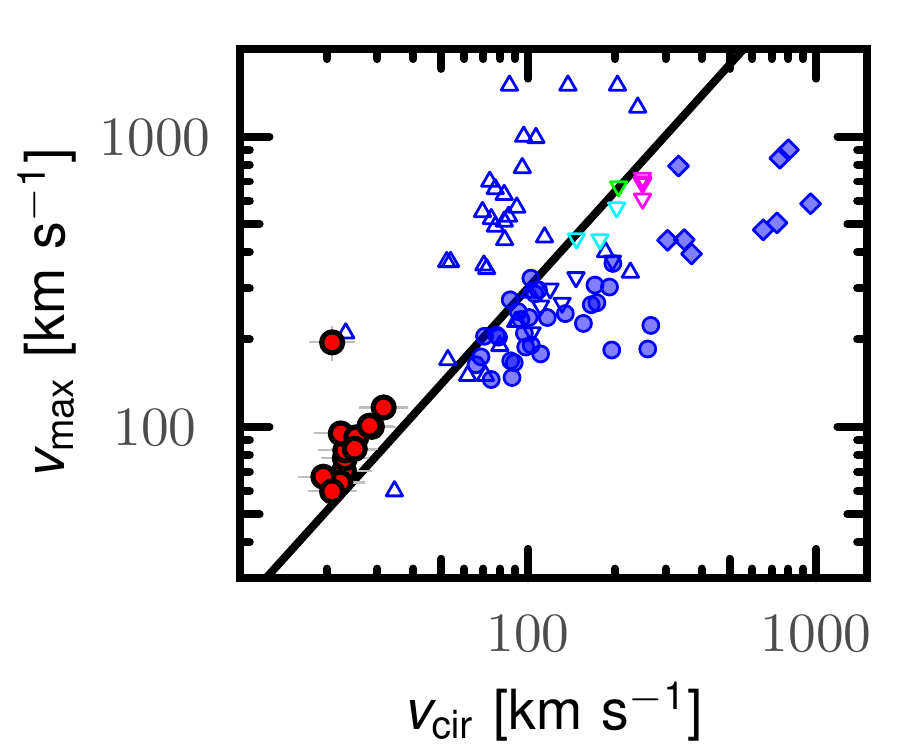}
    \caption{Maximum outflow velocity ($v_\mathrm{max}$) as a function of circular velocity ($v_\mathrm{cir}$). The symbols are the same as Figure \ref{fig:vmax_mass_sfr}. The black line is the $v_\mathrm{max}-v_\mathrm{cir}$ relation given by \cite{Muratov+15}. We shrink the red circles to avoid a busy plot.}
    \label{fig:outflow_vcir}
\end{figure}

\subsection{Scaling Relations of Outflow Velocity and Comparison with Previous Studies}
\label{discussion_vmax}

Outflow velocities are known to scale with host galaxy properties including stellar masses and SFRs \citep[e.g.,][]{Martin+05,Chisholm+15}.
This study explores the scaling relations down to $M_*\sim10^5~M_\odot$.
In the bottom panels of Figure \ref{fig:vmax_mass_sfr}, 
we show the correlations of $v_\mathrm{max}$ with $M_*$ and SFR for our galaxies and galaxies investigated by previous studies.
The same definition of $v_\mathrm{max}$ is adopted by \cite{Bruno+19} and this study.
For galaxies taken from \cite{Perrotta+21}, we calculate $v_\mathrm{max}$ with Equation (\ref{eq:vmax}).
\cite{Heckman+16} and \cite{Sugahara+17,Sugahara+19} estimate $v_\mathrm{max}$ with interstellar UV absorption lines.
We note that, between the outflow velocities measured by emission lines and absorption lines, a comparison is hard to make without knowing the precise velocity distribution in outflowing gas.
We mainly focus on the comparison with previous results that are derived from emission lines.
Most galaxies investigated by previous studies and this study have high SFRs, suggested by the SFR-$M_*$ distribution in Figure \ref{fig:sfr_m}.

In the bottom left panel of Figure \ref{fig:vmax_mass_sfr}, 
a positive correlation between the outflow velocity and the stellar mass is identified for our galaxies, with Spearman rank correlation coefficients of $\rho\sim0.60$ ($\sim0.60$) and $p\sim0.028$ ($\sim0.026$) for the H$\alpha$ ([{\sc Oiii}]) lines.
Down to $M_*\sim10^5~M_\odot$, we obtain outflow velocity as small as $\sim40~\mathrm{km~s^{-1}}$.
In comparison, massive galaxies of $M_*\sim10^{11}~M_\odot$ can host outflow as fast as $\sim1000~\mathrm{km~s^{-1}}$.
The positive correlation $v_\mathrm{max}-M_*$ correlation for our sample is generally consistent with the finds of previous studies that use emission lines.
Our galaxies also show a positive $v_\mathrm{max}-$SFR correlation, as presented in the bottom right panel of Figure \ref{fig:vmax_mass_sfr}.
The Spearman rank correlation coefficients are $\rho\sim0.71$ ($\sim0.78$) and $p\sim0.006$ ($\sim0.002$) for the H$\alpha$ ([{\sc Oiii}]) lines.
\cite{Bruno+19} and \cite{Perrotta+21} conclude no clear $v_\mathrm{max}-$SFR correlation.
Combining all the previous data, it is likely that a positive $v_\mathrm{max}-SFR$ correlation exists over a large SFR range.
Interestingly, our galaxies are offsetted from this correlation.
Our galaxies have $v_\mathrm{max}$ smaller than those obtained by \cite{Bruno+19} by at least a factor of three, despite the same SFRs.
We thus conclude that the scaling relation between $v_\mathrm{max}$ and SFR have large scatters due to different galaxy properties including stellar masses.
This result implies that SFR cannot be treated as the fundamental parameter of determining outflow velocity.

Many studies have investigated the dependence of $v_\mathrm{max}$ on the circular velocity of DM Halo.
\cite{Sugahara+19} discuss the key role of $v_\mathrm{cir}$ in determining both the gravitational potential and star-forming activity over different redshifts.
They conclude that $v_\mathrm{cir}$ is probably the fundamental parameter that determines outflow velocity.
Their results also agree well with the simulation results obtained by \cite{Muratov+15}. 
In Figure \ref{fig:outflow_vcir}, we show the dependence of $v_\mathrm{max}$ on $v_\mathrm{cir}$ and find a positive correlation for our sample and the galaxies taken from \cite{Bruno+19}, \cite{Sugahara+17,Sugahara+19}, and \cite{Heckman+16}.
For data taken from \cite{Bruno+19} and \cite{Heckman+16}, we convert the stellar masses to $v_\mathrm{cir}$ following the procedures taken by \cite{Sugahara+19}.
The solid line represents the simulation results of \cite{Muratov+15}.
Although \cite{Muratov+15} use the 95th percentile outflow velocity at 25\% of virial radius, the differences from definitions of the outflow velocity is within the scatter of observational results. 
In this study, we extend the $v_\mathrm{max}-v_\mathrm{cir}$ relation in \cite{Sugahara+19} and \cite{Muratov+15} down to $v_\mathrm{cir}\sim10~\mathrm{km~s^{-1}}$.

\subsection{Gas Escaping}
Outflows are considered an important mechanism to eject gas into the IGM, especially for low-mass galaxies that have shallow gravitational potentials.
We estimate the escape velocity ($v_\mathrm{esc}$) for our galaxies to examine whether the outflows are fast enough to escape from the gravitational potentials.
Assuming that the DM halo is an isothermal sphere truncated at a radius $r_\mathrm{max}$,
$v_\mathrm{esc}$ at the radius $r$ can be estimated from the circular velocity \citep{Heckman+00}:
\begin{equation}
    v_\mathrm{esc}=v_\mathrm{cir}\{2[1+\mathrm{ln}(r_\mathrm{max}/r)]\}^{\frac{1}{2}}.
\end{equation}
The value of $r_\mathrm{max}$ is approximated by the halo radius calculated in Section \ref{galaxy_property}.
Note that the drag force is ignored in this estimation.
The escape velocity is not sensitive to the value of $r_\mathrm{max}/r$ in the range of $10-100$ \citep{Veilleux05}.
We thus adopt $v_\mathrm{esc}=3v_\mathrm{cir}$, and obtain $v_\mathrm{esc}\sim60-130~\mathrm{km~s^{-1}}$ for our galaxies as listed in Table \ref{tab:outflow}.

\begin{figure}[t!]
    \centering
    \includegraphics[width=\linewidth]{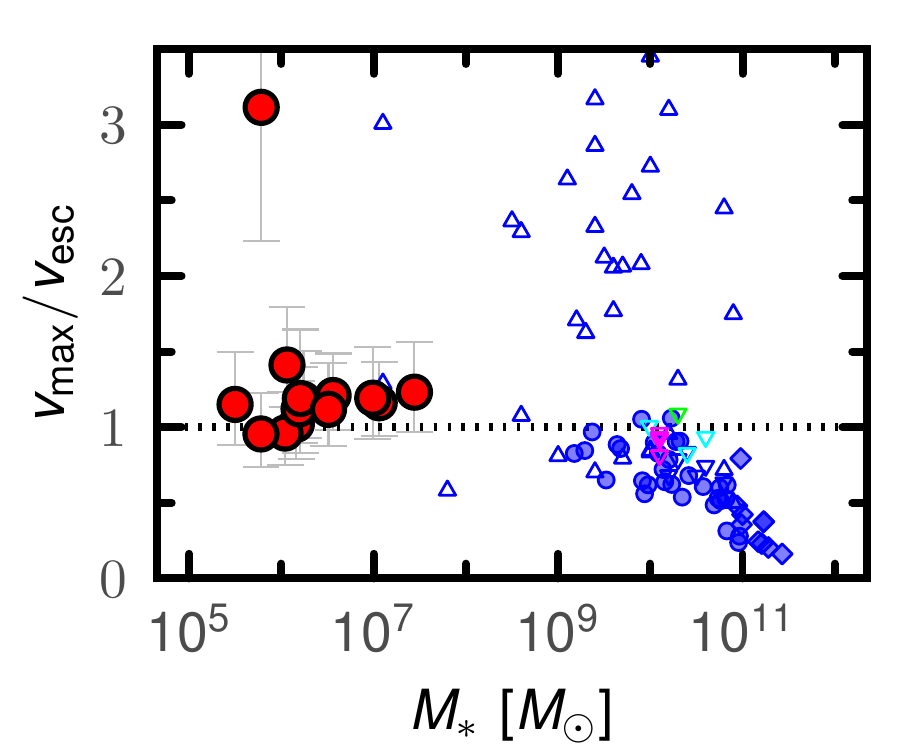}
    \caption{$v_\mathrm{max}/v_\mathrm{esc}$ as a function of stellar mass. The symbols are the same as Figure \ref{fig:vmax_mass_sfr}.} 
    \label{fig:vesc_mass}
\end{figure}

In Figure \ref{fig:vesc_mass}, we show $v_\mathrm{max}/v_\mathrm{esc}$ for our galaxies, where the 1$\sigma$ errors are mainly from the scatter of the $M_*-M_\mathrm{h}$ relation (Section \ref{galaxy_property}).
Most of our galaxies have $v_\mathrm{max}/v_\mathrm{esc}$ slightly larger than unity.
However, we cannot confirm or rule out the possibility for the outflowing gas to escape due to the large uncertainties.
In particular, J2253$+$1116 has $v_\mathrm{max}/v_\mathrm{esc}\sim3$, which is significantly larger than unity.
Given that the H$\alpha$ line in J2253$+$1116 has relatively weak broad components, deeper observations are needed to reveal the fate of outflowing gas in this galaxy.

\cite{Bruno+19} and \cite{Arribas+14} also show $v_\mathrm{max}/v_\mathrm{esc}$ as a function of stellar mass and find that galaxies with smaller stellar masses are more likely to have $v_\mathrm{max}/v_\mathrm{esc}>1$.
By comparing our results with previous ones with emission lines in Figure \ref{fig:vesc_mass},
we do not find a increasing trend of $v_\mathrm{max}/v_\mathrm{esc}$ below $\sim 10^8~M_\odot$.
One the other hand, \cite{Heckman+16} find galaxies with extreme outflows that are fast enough to escape over a large range of stellar masses.
In Figure \ref{fig:vesc_mass}, most galaxies taken from \cite{Heckman+16} have $v_\mathrm{max}/v_\mathrm{esc}>1$.
One galaxy with $M_*\sim 10^7~M_\odot$, Haro 3, shows $v_\mathrm{max}/v_\mathrm{esc}\sim3$ similar to J2253$+$1116.
Despite the differences of $v_\mathrm{max}$ definitions, there likely exist outflows with velocity fast enough to escape in low-mass regime.

We note that ten out of the sixteen galaxies show diffuse tails on the HSC/SDSS images.
This tadpole geometry is commonly seen in local low-mass star-forming galaxies \citep[][]{Morales-Luis+11}.
The diffuse tails can be ten times as massive as the respective galaxies, which is suggested by \cite{Isobe+21}, who estimate the stellar masses of the diffuse tails of J1142$-$0038 and J2314$+$0154.
If our galaxies reside in the same DM halos as the diffuse tails, the halo mass and escape velocity may be underestimated.
Re-calculating escape velocities with stellar masses ten times of the original values, we find that the escape velocities would be $\approx30$ percent larger than the values shown in Table \ref{tab:outflow}.
With these escape velocities, the majority of our galaxies would have $v_\mathrm{max}/v_\mathrm{esc}\lesssim1$.
The conclusion is unchanged given the large uncertainties of $v_\mathrm{esc}$.
Further investigations on the dynamics of low-mass galaxies will be helpful to better constrain the escape of gas.

\begin{figure*}[t]
\centering
\includegraphics[width=0.75\textwidth]{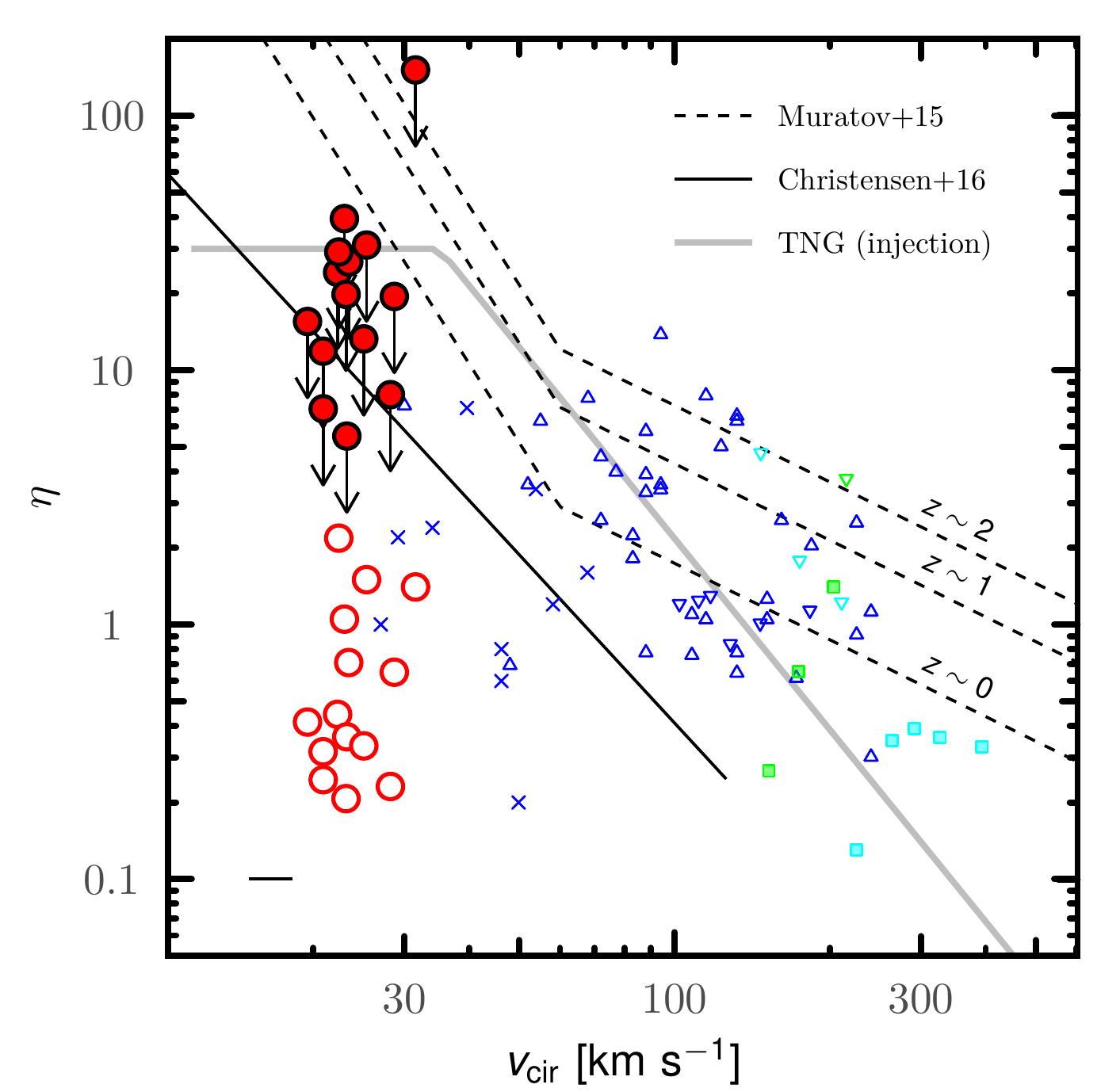}
\caption{Mass loading factor ($\eta$) and its dependence on $v_\mathrm{cir}$. 
We estimate $\eta$ with fiducial parameters and extreme parameters, shown as the open and close solid circles, respectively, (see text). The solid line in the left-bottom corner indicates the typical error of $v_\mathrm{cir}$ for our galaxies. We include the predictions of two `zoom' simulations \citep[][]{Muratov+15,Christensen+18} and the prescription used at injection by IllustrisTNG \citep[e.g.,][]{Pillepich+18,Nelson+19}.} The other symbols are the same as those in Figure \ref{fig:BN_mass_sfr} and \ref{fig:vmax_mass_sfr}.
\label{fig:eta_vcir}
\end{figure*}

\subsection{Mass Outflow and Implications for Feedback}
\label{eta}
The ratio of outflow mass loss rate ($\dot{M}_\mathrm{out}$) to star-formation rate
is referred to as the mass-loading factor ($\eta$):
\begin{equation}
\label{eq:eta}
    \eta = \frac{\dot{M}_\mathrm{out}}{\mathrm{SFR}}.
\end{equation}
This quantity is widely used by observations \citep[e.g.,][]{Heckman+15,Freeman+19,Sugahara+17} and simulation studies \citep[e.g.,][]{Muratov+15}.
Following the procedures taken by \cite{Newman+12} and \cite{Freeman+19}, we calculate the mass of outflowing gas using the equation:
\begin{equation}
\label{eq:Mout}
    M_\mathrm{out}=\frac{1.36m_\mathrm{p}
    L_\mathrm{H\alpha,b}}{\gamma_\mathrm{H\alpha}n_\mathrm{e}},
\end{equation}
where $L_\mathrm{H\alpha,b}$ is the extinction-corrected H$\alpha$ luminosity of the broad component.
The quantity $m_\mathrm{p}$ is the atomic mass of hydrogen and $n_\mathrm{e}$ is the electron density in the outflowing gas.
For the volume emissivity of H$\alpha$ ($\gamma_\mathrm{H\alpha}$), we use $\gamma_\mathrm{H\alpha}=3.56\times10^{-25}~\mathrm{erg~cm^{-3}~s^{-1}}$, assuming the case B recombination and an electron temperature of $T_\mathrm{e}=10^4~\mathrm{K}$.

We assume that the outflowing gas moves in a solid angle $\Omega$ with a radially constant velocity $v_\mathrm{max}$ and calculate the mass outflow rate with:
\begin{equation}
\label{eq:dotMout}
    \dot{M}_\mathrm{out}=M_\mathrm{out}\frac{v_\mathrm{max}}{r_\mathrm{out}},
\end{equation}
where $r_\mathrm{out}$ is the radial extent of outflow.
Combining Equations (\ref{eq:eta})-(\ref{eq:dotMout}) and adopting Equation (\ref{eq:sfr}) for SFR, 
we obtain the following calculation for $\eta$:
\begin{align}
    \eta
    &\approx0.75\left(\frac{v_\mathrm{max}}{n_\mathrm{e}r_\mathrm{out}}\right)
    \left(\frac{L_\mathrm{H\alpha,b}}{L_\mathrm{H\alpha,b}+L_\mathrm{H\alpha,n}}\right) \\
    &\approx0.75\left(\frac{v_\mathrm{max}}{n_\mathrm{e}r_\mathrm{out}}\right)
    \left(\frac{BNR}{1+BNR}\right),
\label{eq:eta_final}
\end{align}
where $L_\mathrm{H\alpha,n}$ is the extinction-corrected H$\alpha$ luminosity of the narrow component.
The units of $v_\mathrm{max}$, $n_\mathrm{e}$, and $r_\mathrm{out}$ are $\mathrm{km~s^{-1}}$, $\mathrm{cm^{-3}}$, and $\mathrm{kpc}$, respectively.
In Equation (\ref{eq:eta_final}), assuming that the emission of narrow and broad components have the same dust attenuation, we approximate the ratio of extinction-corrected luminosity using the flux ratio (BNR) of the H$\alpha$ line.
We note that we do not multiply Equation (\ref{eq:eta_final}) by a factor of two as in \cite{Newman+12} because we do not find that the the red wings of our broad components are significantly obscured.
Similar to previous studies, our measurements of $\eta$ are subject to the large uncertainties given by the estimation of $n_\mathrm{e}$ and $r_\mathrm{out}$.
Therefore we present two estimations of $\eta$: fiducial values calculated with $r_\mathrm{out}$ and $n_\mathrm{e}$ that our low-mass galaxies likely have,
and maximum values ($\eta_\mathrm{max}$) calculated with small $r_\mathrm{out}$ and $n_\mathrm{e}$ that produce relatively large $\eta$ in extreme cases.

Electron density of the outflowing gas can be estimated if emission line doublets like [{\sc Oii}]$\lambda\lambda3727,3729$ and [{\sc Sii}]$\lambda\lambda6716,6731$ are decomposed into narrow and broad components, where broad components originate from outflows.
\cite{Arribas+14} and \cite{Bruno+19} obtain median values of
$n_\mathrm{e}\sim315~\mathrm{cm}^{-3}$ and $302~\mathrm{cm}^{-3}$, respectively, for the broad components of [{\sc Sii}] doublets.
Both of \cite{Arribas+14} and \cite{Bruno+19} find that electron densities measured from the broad components are larger than those from the narrow components.
There are two models of outflowing gas proposed by previous studies \citep[e.g.,][]{Genzel+11,Newman+12}, 
one of which can explain the observational results of \cite{Arribas+14} and \cite{Bruno+19},
assuming that outflowing gas forms compact clouds and retains relatively high local electron density.
However, electron density decreases with radius in the other model that assumes outflowing gas filling the entire volume of outflow cone.
In this study, we do not have enough SNRs to measure the line ratios of the broad components in [{\sc Oii}] and [{\sc Sii}] doublets, 
while the systemic values of electron density can be measured from [{\sc Oii}] doublets with single-Gaussian fitting.
We therefore adopt the systemic values as the fiducial values of $n_\mathrm{e}$ in Equation (\ref{eq:eta_final}), given that we do not know which model better explains the outflowing gas in our galaxies.
For galaxies without measurements of electron density, we assume a typical value of $100~\mathrm{cm^{-3}}$.
We use $n_\mathrm{e}\sim10~\mathrm{cm^{-3}}$ to evaluate $\eta_\mathrm{max}$.

The value of $r_\mathrm{out}$ could be the main source of uncertainties in the estimation of $\eta$ using the emission line method.
The observational measurements of $r_\mathrm{out}$ is difficult and sensitive to the assumption of outflow geometry.
\cite{McQuinn+19} measure the extent of the H$\alpha$ emission in local dwarf galaxies with narrow band images.
Results of \cite{McQuinn+19} show that the diffused ionized gas reach the projected distances of $\sim1-5~\mathrm{kpc}$, that are $\sim2-3$ times the radius of the H{\sc i} disk.
They estimate $\eta$ based on the amount of gas that pass through a thin shell whose thickness is equivalent to our definition of $r_\mathrm{out}$. 
\cite{Newman+12} utilize the integral field spectra of a $z\sim2$ clumpy star-forming galaxy and approximate $r_\mathrm{out}$ with the radius of the star-forming clumps.
In their case, the mass of outflowing gas is calculated from the flux of the broad emission line components.
Motivated by the method of \cite{Newman+12}, we adopt the effective radius ($r_\mathrm{e}$) as the our estimations for $r_\mathrm{out}$.
For 15 galaxies selected by the EMPRESS project, \cite{Isobe+21} obtain $r_\mathrm{e}\sim40-2500~\mathrm{pc}$ and find a size-mass relation similar to the $z\sim0$ star-forming galaxies in \cite{Shibuya+15}.
We adopt the median (minimum) value of $r_\mathrm{e}\sim150~\mathrm{pc}$ ($40~\mathrm{pc}$) as the fiducial (extreme) 
$r_\mathrm{out}$ values for the 14 galaxies in this paper.
The values of $r_\mathrm{e}$ measured by \cite{Isobe+21} has a standard deviation of $\sim600~\mathrm{pc}$, which is larger than the range of $r_\mathrm{e}$ given by the size-mass relation between $M_*\sim10^5-10^7~M_\odot$.
Therefore, we cannot improve $r_\mathrm{e}$ estimations by applying the size-mass relation.
% the range of $r_\mathrm{e}$ is smaller than $\sim600~\mathrm{pc}$, which means $r_\mathrm{out}$ is not better constrained.
% If we apply the size-mass relation to obtain rough estimations of $r_\mathrm{out}$ for each galaxy, the uncertainties of $\eta$ originated from the size-mass relation would be a factor of 4.
% Because a factor of 4 is much smaller than the differences between $\eta$ and $\eta_\mathrm{max}$,
Because we discuss the uncertainties originated from $r_\mathrm{out}$ by calculating $\eta$ and $\eta_\mathrm{max}$, we apply the same $r_\mathrm{out}$ value for all galaxies.

Equation (\ref{eq:dotMout}) is applicable for expanding shells or spheres as pointed out in \cite{Olmo-Garcia+17}, if an appropriate extent of outflows is chosen.
It is also used in studies that assume biconical outflows.
Here we test different geometries of outflows by evaluating $r_\mathrm{out}$ assuming a sphere filled with ionized gas that has a density of $1.36m_\mathrm{p}n_\mathrm{e}$:
\begin{equation}
    r_\mathrm{out,sphere}=\left(\frac{3}{4\pi}\frac{M_\mathrm{out}}{1.36m_\mathrm{p}n_\mathrm{e}}\right)^{1/3}.
    \label{eq:rout_sphere}
\end{equation}
We obtain $r_\mathrm{out,sphere}\sim10-80~\mathrm{pc}$ combining Equations (\ref{eq:Mout}) and (\ref{eq:rout_sphere}).
Our choice of the fiducial and extreme values of $r_\mathrm{out}$ are of the same order as $r_\mathrm{out,sphere}$.
If the outflowing gas moves as an expanding shell, the extent of outflows would be smaller as the radius of the shell increases ($r_\mathrm{out} < r_\mathrm{out,sphere}$).
However, if the outflowing gas has a clumpy structure or a non-spherical geometry (e.g., biconical outflows), an extent larger than $r_\mathrm{out,sphere}$ needs to be assumed ($r_\mathrm{out} > r_\mathrm{out,sphere}$).
We show in Section \ref{profile_discuss} that the broad components are likely originated from a spherical or conical structure filled by ionized gas instead of a thin shell.
Therefore, outflows in our galaxies are probably characterized by our choice of $r_\mathrm{out}$.

Finally, we obtain $\eta$ ($\eta_\mathrm{max}$) values of $0.21-2.18$ ($4.78-131.39$) with a median value of $0.43$ ($17.09$).
We present the results in Table \ref{tab:outflow}, and show $\eta$ and $\eta_\mathrm{max}$ as a function of $v_\mathrm{cir}$ in Figure \ref{fig:eta_vcir}.
Galaxies taken from \cite{Heckman+15}, \cite{Freeman+19}, \cite{McQuinn+19}, \cite{Sugahara+17}, and \cite{Swinbank+19} are compared in Figure \ref{fig:eta_vcir}.
We estimate the value of $v_\mathrm{cir}$ from halo mass as explained in Section \ref{discussion_vmax}.
\cite{McQuinn+19} investigate $\eta$ as a function of $v_\mathrm{cir}$ that is defined for the H{\sc i} region. 
To make a consistent comparison, here we use the stellar masses of their galaxies to evaluate $v_\mathrm{cir}$ that correspond to circular velocity on the virial radius of dark matter halo.
Our estimations of $\eta$ are comparable with those of high mass galaxies with $v_\mathrm{cir}\sim50-300~\mathrm{km~s^{-1}}$ within one order of magnitude.

The black dashed lines in Figure \ref{fig:eta_vcir} are the predictions of the Feedback in Realistic Galaxies (FIRE) simulations conducted by \cite{Muratov+15}, who find a boundary of $v_\mathrm{cir}=60~\mathrm{km~s^{-1}}$ below which $\eta$ increase steeply towards small $v_\mathrm{cir}$ ($\eta\propto v_\mathrm{cir}^{-3.2}$).
The steep slope below $v_\mathrm{cir}=60~\mathrm{km~s^{-1}}$ is originated from the simulation results of low-mass galaxies.
Among the $z\sim0$ simulations in \cite{Muratov+15}, the least massive halo with $v_\mathrm{cir}\sim33~\mathrm{km~s^{-1}}$ has $\eta\sim45$.
% One interpretation for their results is that outflows are momentum-driven in the high $v_\mathrm{cir}$ galaxies and energy-driven in the low $v_\mathrm{cir}$ galaxies.
\cite{Christensen+18} also explore low-mass galaxies with `zoom' simulations and predict a $\eta\propto v_\mathrm{cir}^{-2.2}$ relation as shown with the black solid line in Figure \ref{fig:eta_vcir}.
The slope of $-2.2$ given by \cite{Christensen+18} is close to the slope of $-2$ for energy-driven outflows but flatter than the one predicted by \cite{Muratov+15}.
% The difference between the $\eta$ calculated by \cite{Muratov+15} and \cite{Christensen+18} can possibly be explained by different definitions of outflows.}
Although our galaxies reside in the regime of energy-driven outflows, 
% our results do not show large $\eta$ values predicted by simulations, 
our estimation of $\eta$ is smaller than the predictions of \cite{Muratov+15} and \cite{Christensen+18}.
For $\eta_\mathrm{max}$ that is calculated with extreme parameters, our results are consistent with the predictions of \cite{Christensen+18} but still smaller than those of \cite{Muratov+15}.
One exception is J1253$-$0312 that has $\eta_\mathrm{max}\sim131.39$, suggesting strong feedback.
However, we emphasize the choice of $r_\mathrm{out}\sim40~\mathrm{pc}$ and $n_\mathrm{e}\sim10~\mathrm{cm^{-3}}$ is extremely low, especially for J1253$-$0312 that has the largest stellar mass in our galaxies.
In other words, even in the case that outflows have $v_\mathrm{max}\sim150~\mathrm{km~s^{-1}}$ and BNR$\sim2$, we still need to assume concentrated geometry and low electron density to produce large $\eta$.
On the other hand, the fiducial $\eta$ values are significantly below the simulation predictions, possibly suggestive of the feedback being weak in our galaxies.
As discussed in Section \ref{results}, the detection of weak outflows is possible because our spectra have sufficiently high SNRs.
We also compare our results with the prescription used by The Next Generation Illustris \citep[IllustrisTNG,][]{Pillepich+18}.
In the low-mass regime, IllustrisTNG assumes a minimum wind velocity at injection and adopts a maximum mass loading factor that is larger than our estimations.
However, the outflows in low-mass regime are still unexplored by large-volume simulations like TNG50 \citep[][]{Nelson+19}.
Therefore, adopting a prescription with weak feedback ($\eta\sim1$) is potentially interesting for future simulations that resolve low-mass galaxies.

We note that different phases of outflowing gas are used by simulations and our methods in the calculation of $\eta$.
% Observational studies \citep[e.g.,][]{Veilleux05} show that the outflows are composed of hot ($T\sim10^6~\mathrm{K}$), warm ($T\sim10^4~\mathrm{K}$), and cold phases ($T\sim10^1-10^3~\mathrm{K}$).
The H$\alpha$ and [{\sc Oiii}] emission is originated from the warm gas.
Therefore the $\eta$ values derived from emission line methods are usually treated as lower limits \citep{Newman+12} because a significant fraction of outflowing mass can be in cold atomic or molecular phase that is not detected with the emission line methods.
Simulations including \cite{Muratov+15}, on the other hand, usually include gas in all phases.
For our galaxies, active star formation may produce strong ionization radiation that ionizes most of the diffused gas.
With MUSE observations, \cite{Bik+18} find a largely extended H$\alpha$ halo around a low-mass galaxies, ESO 338, that has similar properties to our galaxies,
suggestive that the most gas is in warm ionized phase with the strong ionization radiation from the center.
For another similar galaxy, Haro 11, observed with MUSE, 
\cite{Menacho+19} find that the neutral atomic gas mass is one-third of the ionized gas mass, while the molecular gas mass could be comparable to the ionized gas.
It is likely that the mass contribution from cold phase outflows are relatively small compared to warm phase outflows in local compact star-forming galaxies.
If we adopt the mass ratio in \cite{Menacho+19},
including the cold-phase outflows may only increase our $\eta$ estimations by a factor of two, which does not impact our conclusions.

\subsection{Other Interpretations of the Broad Components}
\label{profile_discuss}

\begin{figure}[t!]
\centering
\includegraphics[width=\linewidth]{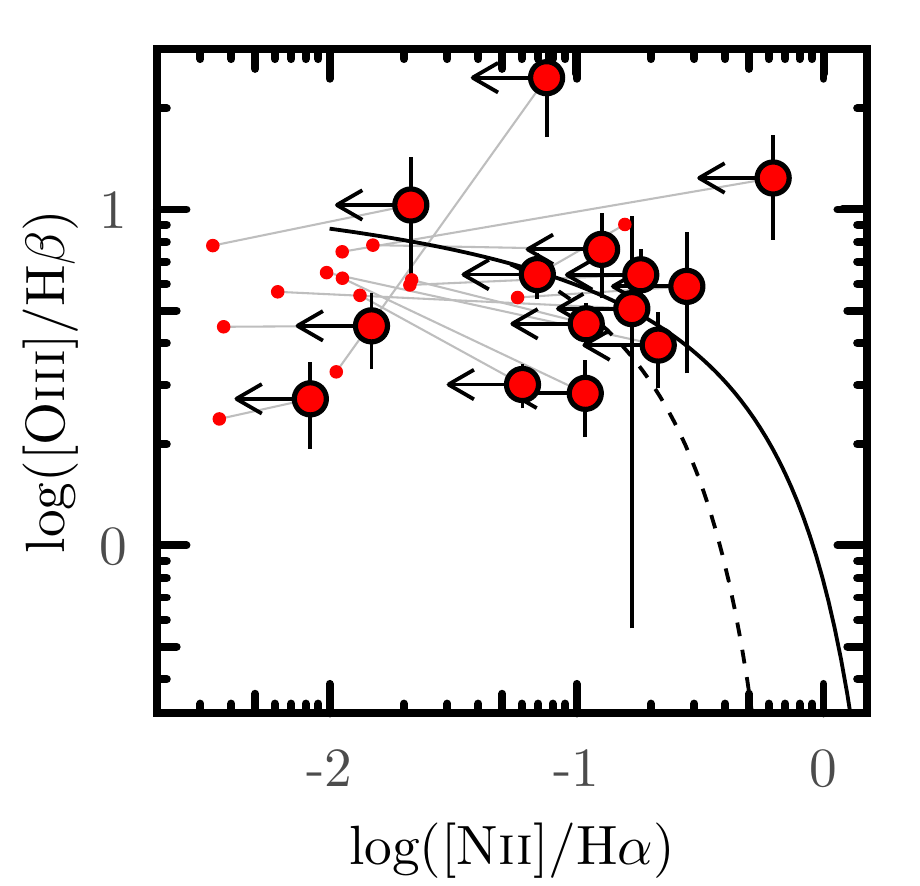}
\caption{BPT diagnostics diagram. The small red dots are derived from the narrow components of the emission lines, while the red circles from the broad components. In particular, the [{\sc Nii}]$\lambda6583$ lines are not fit with the double-Gaussian profile, therefore the total fluxes are used, see text. The solid and dashed lines are taken from \cite{Kewley+01} and \cite{Kauffmann+03}, respectively.}
\label{fig:BPT}
\end{figure}

Profile fitting with multiple components is commonly used to probe outflows.
However, alternative interpretations of the secondary components can be proposed \citep[e.g.,][]{Olmo-Garcia+17, Freeman+19}. 
Here we discuss different possibilities that are rotation of galaxies, accretion disks around massive objects (e.g., BHs), gas inflows, and expanding bubbles.

In this study, we successfully decompose the emission lines with double-Gaussian profiles, showing not only the existence of two components, but also clear deviation between the line widths of the two components, with $\sigma_\mathrm{b}-\sigma_\mathrm{n} \sim20-100~\mathrm{km~s^{-1}}$.
According to the Tully-Fisher relation \citep[][]{TF}, galaxies with $M_*\sim10^6-10^7~M_\odot$ are characterized by rotation velocities of $\lesssim30~\mathrm{km~s^{-1}}$, while we obtain $\mathrm{FWHM_b}\gtrsim100~\mathrm{km~s^{-1}}$.
Even we assume that the gas contributes to 99\% of the baryonic mass and refer to the baryonic Tully-Fisher relation \citep[][]{McGaugh+00}, the typical rotation velocity for a galaxy with a baryonic disk mass of $\sim10^9~M_\odot$ is $\lesssim100~\mathrm{km~s^{-1}}$.
Therefore, the broad components most likely represent gas components with relatively high velocities (e.g., outflows) that are distinctive from the rotation of galaxies.

The accretion disk around massive objects can produce broad emission lines.
In our galaxies, the fluxes from the broad components are of the same order of magnitude as those from the narrow components ($BNR\sim0.3$).
The broad line region (BLR) around a supermassive BH can emit strong emission lines that are broadened by the rotational motion of the accretion disk, with $\mathrm{FWHM}\gtrsim1000~\mathrm{km~s^{-1}}$.
However, the FWHMs of the broad components in this study are significantly smaller than those found in BLRs.
% By assuming a radius of $1~\mathrm{pc}$ for the rotation disk, we can obtain rough estimates for the mass of the central objects.
% We use the FWHMs of the H$\alpha$ lines and obtain $10^6-10^7~M_\odot$ for the central objects, which is not practical compared to the stellar masses of our galaxies.
% For the accretion disks, broad lines only form in the high density gas clouds located close to the central objects.
The BLRs and the accretion disks are also known to have high densities, which results in the absence of forbidden lines.
As we pointed out in Section \ref{results}, the prominent broad components in the [{\sc Oiii}] are only produced in ionized gas with low density.
Therefore, the broad components found in this study do not characterize BLRs.

We further check the excitation states for the gas traced with different components using the well-known BPT diagnostic diagram.
We conduct double-Gaussian fitting to the H$\beta$ lines following the procedures in Section \ref{fitting1d}.
For 13 out of the 14 galaxies, we successfully obtain best-fit double-Gaussian profiles and derive the line fluxes for the narrow and broad components.
For J1418$+$2102 whose H$\beta$ line is too faint, we use single-Gaussian fitting and derive the line fluxes of the narrow and broad components adopting the $BNR$ value of the H$\alpha$ line.
The [{\sc Nii}]$\lambda6583$ lines are too faint to be reliably fitted with the double-Gaussian profiles, therefore we only use the the line fluxes derived from the best-fit single-Gaussian profiles.
In other words, we derive the upper limits of [{\sc Nii}]/H$\alpha$, assuming one single component contributes to all the [{\sc Nii}] emission.
In Figure \ref{fig:BPT}, we show the BPT diagram for the narrow and broad components.
Except for J1253$-$0312, narrow components for 13 out of the 14 galaxies fall in the region of star-forming galaxy.
For the broad components, 12 out of the 14 galaxies locate in the region of star-forming galaxy or close to the classification line \citep[][]{Kewley+01}.
The other two objects J2253$+$1116 % (highest N2/Ha)
and J1401$-$0040  % (highest O3/Hb)
are offset from the star-forming region, indicating possible excitation from active galactic nuclei (AGN) or fast radiative shock.
We note again that the [{\sc Nii}]/H$\alpha$ of the broad components are uppler limits, that can be lowered by a factor of $\sim3$ if we adopt a typical value of $BNR=0.3$.
Therefore, there is no clear evidence showing the connection between the broad components and AGNs.
Similarly, the possibility of inflows can be ruled out given that the gas is highly ionized ([{\sc Oiii}]/H$\beta > 1$), while the inflow is expected to be composed of cold gas.

An expanding bubble is produced when sufficient energy or momentum is injected into the ISM. 
When the expanding bubble reaches the disk scale, the bubble breaks up and possibly create outflows with a conical geometry.
By observations, most ionized outflowing gas moves as a thin shell or along the walls of conical structures \citep[see][and references therein]{Veilleux05}.
In the case of the expanding bubble, an optically thin, symmetric thin shell, would produce emission lines with a top-hat profile \citep[e.g.,][]{Cid+94}.
\cite{Olmo-Garcia+17} observe ``double-horn'' structures in the line profiles that are pairs of secondary components on both sides of the emission lines.
They discuss different origins of emission line profiles and conclude that the ``double-horn'' structure likely emerges from a shell with dust extinction.
In our case, the broad wings of an emission line cannot be explained by the ``double-horn'' structure but one broad component with a Gaussian shape.
The inner part of a Gaussian shape represents the gas with small line-of-sight velocities.
Therefore, the Gaussian shape indicates that the gas has a range of velocities, which favors the explanation that outflowing gas is moving at different radii instead of on one thin shell.
Therefore, a thin shell structure is unlikely for the broad components we detect, while either a spherical or conical morphology can be adopted in our analysis.

\section{Summary}
\label{summary}
We analyze the profiles of the H$\alpha$ and [{\sc Oiii}] emission lines in 21 nearby low-mass galaxies
with masses of $M_*\sim10^4-10^7~M_\odot$.
Spectra of thirteen galaxies were newly taken with Magellan/MagE.
We find evidence for warm ionized gas outflows in 14 out of the 21 galaxies, and study the outflow properties in the low-mass regime.
Our findings are summarized below.

\begin{enumerate}
    \item For our galaxies, we do not find a clear correlation between $BNR$ and $M_*$ (SFR). However, we obtain mean values of $BNR\sim0.31$ and 0.21 for the H$\alpha$ and [{\sc Oiii}] lines, respectively, that are generally smaller than those of massive galaxies.
    The relatively high SNR of our spectra may allow us to detect the smaller $BNR$ values characteristic of weaker outflows.
    \item We find strong evidence of smaller $v_\mathrm{max}$ towards lower $M_*$ and SFR.
    Combing our sample with existing data from previous studies, we confirm a positive correlation between $v_\mathrm{max}$ and SFR, but find a large scatter of $v_\mathrm{max}$ for a given SFR.
    We also explore the $v_\mathrm{max}-v_\mathrm{cir}$ relation below $v_\mathrm{cir}\sim30~\mathrm{km~s^{-1}}$, showing consistent results with previous observations for massive galaxies \citep[e.g.,][]{Sugahara+19} and the predictions from the simulations by \cite{Muratov+15}.
    \item We investigate whether the outflowing gas is fast enough to escape from the galaxies by estimating the escape velocities. However, we cannot conclude whether the outflowing gas can escape in most of our galaxies due to the large uncertainties given by $v_\mathrm{esc}$ estimation.
    \item We evaluate the fiducial values of mass-loading factors $\eta$ with estimated $n_\mathrm{e}$ and $r_\mathrm{out}$.
    We also provide maximum $\eta$ values with extreme parameters.
    Our results point to relatively weak stellar feedback in our galaxies.
    Even if we choose extreme parameters, we find $\eta$ values are generally smaller than those predicted by simulations of low-mass galaxies. 
\end{enumerate}

\begin{acknowledgements}
We thank the anonymous referee for constructive comments and suggestions.
We are grateful to Kazuhiro Shimasaku, Ricardo O. Amor\'{i}n, Xinfeng Xu, Alejandro Lumbreras-Calle, Dylan Nelson for their useful comments and discussions. 

This work is supported by the World Premier International
Research Center Initiative (WPI Initiative), MEXT, Japan, as
well as KAKENHI Grant-in-Aid for Scientific Research (A)
(20H00180 and 21H04467) through the Japan
Society for the Promotion of Science (JSPS).
This work is supported by the joint research program of the Institute for Cosmic Ray Research (ICRR), the University of Tokyo.

This paper includes data gathered with the 6.5 m Magellan Telescopes located at Las Campanas Observatory, Chile. We are grateful to the observatory personnel for help with the observations.
\end{acknowledgements}

\bibliography{ref}{}
\bibliographystyle{aasjournal}

\end{document}